\documentclass[english,american]{iopart}
\usepackage[T1]{fontenc}
\usepackage[latin9]{inputenc}
\usepackage{babel}
\usepackage{prettyref}
\usepackage{float}
\usepackage{amsbsy}
\usepackage{amstext}
\usepackage{feyn}
\usepackage[unicode=true,pdfusetitle,
 bookmarks=true,bookmarksnumbered=false,bookmarksopen=false,
 breaklinks=false,pdfborder={0 0 1},backref=false,colorlinks=false]
 {hyperref}
\hypersetup{
 colorlinks=true,linkcolor=blue,citecolor=blue,urlcolor=blue,filecolor=blue}

\makeatletter
\usepackage{iopams}
\usepackage{setstack}

\usepackage[numbers, sort&compress]{natbib}
\usepackage{iopams}
\usepackage{bbold}

\usepackage{feynmp-auto}
\usepackage{stackrel}

\newrefformat{cap}{\hyperref[#1]{Figure~\ref{#1}}}
\newrefformat{fig}{\hyperref[#1]{Figure~\ref{#1}}}
\newrefformat{tab}{\hyperref[#1]{Table ~\ref{#1}}}
\newrefformat{sec}{\hyperref[#1]{Section~\ref{#1}}}
\newrefformat{sub}{\hyperref[#1]{Section~\ref{#1}}}
\newrefformat{cha}{\hyperref[#1]{Chapter~\ref{#1}}}
\newrefformat{app}{\hyperref[#1]{Appendix~\ref{#1}}}

\makeatletter
\newsavebox{\@brx}
\newcommand{\llangle}[1][]{\savebox{\@brx}{\(\m@th{#1\langle}\)}%
  \mathopen{\copy\@brx\kern-0.5\wd\@brx\usebox{\@brx}}}
\newcommand{\rrangle}[1][]{\savebox{\@brx}{\(\m@th{#1\rangle}\)}%
  \mathclose{\copy\@brx\kern-0.5\wd\@brx\usebox{\@brx}}}
\makeatother

\makeatother

\begin{document}
\global\long\def\T{\mathrm{T}}

\global\long\def\N{\mathcal{N}}
\global\long\def\Z{\mathcal{Z}}
\global\long\def\E{\mathcal{E}}
\global\long\def\C{\mathbb{C}}
\global\long\def\R{\mathbb{R}}
\global\long\def\SR{S_{\mathrm{R}}}
\global\long\def\ST{S_{\mathrm{TM}}}
\global\long\def\SdC{S_{\mathrm{dC}}}

\title{Expansion of the effective action around non-Gaussian theories}

\author{Tobias K\"uhn}

\address{Institute of Neuroscience and Medicine (INM-6) and Institute for
Advanced Simulation (IAS-6) and JARA BRAIN Institute I, Jülich Research
Centre, Jülich, Germany}

\author{Moritz Helias}

\address{Institute of Neuroscience and Medicine (INM-6) and Institute for
Advanced Simulation (IAS-6) and JARA BRAIN Institute I, Jülich Research
Centre, Jülich, Germany}

\address{Department of Physics, Faculty 1, RWTH Aachen University, Aachen,
Germany}
\begin{abstract}
This paper derives the Feynman rules for the diagrammatic perturbation
expansion of the effective action around an arbitrary solvable problem.
The perturbation expansion around a Gaussian theory is well known
and composed of one-line irreducible diagrams only. For the expansions
around an arbitrary, non-Gaussian problem, we show that a more general
class of irreducible diagrams remains in addition to a second set
of diagrams that has no analogue in the Gaussian case. The effective
action is central to field theory, in particular to the study of phase
transitions, symmetry breaking, effective equations of motion, and
renormalization. We exemplify the method on the Ising model, where
the effective action amounts to the Gibbs free energy, recovering
the Thouless-Anderson-Palmer mean-field theory in a fully diagrammatic
derivation. Higher order corrections follow with only minimal effort
compared to existing techniques. Our results show further that the
Plefka expansion and the high-temperature expansion are special cases
of the general formalism presented here.
\end{abstract}
\maketitle

\section{Introduction\label{sec:Introduction}}

Many-particle systems are of interest in various fields of physics.
Field theory offers a versatile unique language to describe such systems
and powerful methods to treat diverse problems arising in classical
statistical mechanics, quantum mechanics, quantum statistics, quantum
field theory, and stochastic dynamical systems \cite{NegeleOrland98,ZinnJustin96,Amit84,Altland01,Vasiliev98}.
Diagrammatic techniques in particular are convenient and efficient
to organize practical calculations that arise in the context of systematic
perturbation expansions and fluctuation expansions around a solvable
problem. But most techniques rely on the solvable part being Gaussian:
The basic connecting elements in Feynman diagrams, or Mayer graphs
\cite{Mayer77}, are lines. The purpose of this paper is to extend
the diagrammatic computation of a central quantity, the effective
action $\Gamma$, beyond this Gaussian case.

The effective action, vertex generating functional, or Gibbs free
energy, is the first Legendre transform of the generating functional
of connected Green's functions or cumulants. Originally developed
in statistical mechanics to study non-ideal gases in the 1930s, its
modern formulation as a Legendre transform was introduced in the context
of quantum statistics \cite{DeDominicis64_14}, but has soon been
applied in quantum field theory \cite{JonaLasinio64_1790} (similar
concepts had been developed in parallel in quantum field theory \cite{Goldstone62}).

The effective action is central to the study of phase transitions,
because its stationary points determine the values of the order parameters
that characterize the states in which matter may exist; it therefore
reduces the problem of a phase transition to finding bifurcation points
in a variational problem \cite{Amit84,ZinnJustin96,Vasiliev98}.
This formulation yields self-consistency equations for the mean value
of a field that incorporate fluctuation corrections. At lowest order,
at the tree level of diagrams, one obtains the mean-field theory of
the Curie-Weiss type. Higher Legendre transforms generalize this concept
to finding self-consistent solutions for renormalized higher order
Green's functions \cite{DeDominicis64_14}\cite[chapter 6]{Vasiliev98}\cite{Berges02_223,Senechal04};
the second transform leading, at lowest order, to a self-consistency
equation for the propagator, the Hartree-Fock approximation \cite{Fock30}\cite[i.p. chapter 6 for a review]{Vasiliev98}.

The second reason for the importance of the effective action is that
it compactly encodes all dynamical or statistical properties of a
system: Connected Green's functions decompose into tree diagrams,
whose vertices are formed by derivatives of $\Gamma$. Conversely,
relatively fewer diagrams contribute to $\Gamma$, making its computation
favorable: Its perturbative expansion around a Gaussian solvable theory,
or around Gaussian fluctuations with regard to a background field,
requires only one-line (particle) irreducible graphs (1PI) \cite{DeDominicis64_14,Amit84,Kleinert09,ZinnJustin96}.

Third, treating critical points by renormalization group methods in
classical \cite{Wilson74_75,Wilson75_773} and quantum systems \cite{Hertz76_1165}
leads to the study of flow equations for the effective action; most
explicitly shown in its functional formulation \cite{WETTERICH93_90,Berges02_223,Metzner12_299}.

Perturbative calculations are essential for most applied problems.
But practically all techniques in field theory are based on expansions
around the Gaussian case. The topic of this article is a systematic,
diagrammatic, and perturbative expansion of the effective action in
the general case, where the solvable problem is not of Gaussian type;
in statistics: the unperturbed problem has cumulants of orders three
or higher; in field theory: the bare theory has non-vanishing connected
Green's functions of orders three or higher; in diagrams: we have
bare propagators that tie together three or more field points instead
of or besides the usual propagator lines.

It may be obvious that the linked cluster theorem in this case remains
intact without change, shown below for completeness. Also the converse
problem, the decomposition of Green's functions into connected Green's
functions by Wick's theorem is immediately replaced by its general
counterpart, the factorization of moments into all product sets of
cumulants \cite{Gardiner85}. But diagrammatic rules for the direct
expansion of the effective action do not follow as simply. Only for
particular non-Gaussian cases, diagrammatic rules have been derived
earlier: for the classical non-ideal gas \cite{Mayer77,Uhlenbeck63}
and for the Ising model \cite{Vasiliev74,Vasiliev75} (compare \cite[end of chap. 6.3.1.]{Vasiliev98}).
For the classical gas, one finds that only so called star-graphs contribute
\cite[chap. 6.3.2.]{Mayer77,Vasiliev74,Vasiliev98}, which is known
as Mayer's second theorem. Star-graphs are diagrams that do not fall
apart by removing one cumulant. For the Ising model one has to introduce
a special notation for the sums, the so called ${\cal N}$-form, which
only includes contributions from star graphs\cite{Vasiliev75}. At
the end of \prettyref{sub:Application-to-Ising}, we will give more
details on this result and show its relation to our work. 

We first derive a recursive diagrammatic algorithm to compute the
effective action perturbatively around a non-Gaussian solvable theory.
Then we show that the correction terms produced by the iteration satisfy
a set of Feynman rules: Only two topologically defined classes of
diagrams remain. The first being irreducible diagrams in a more general
sense than in the known Gaussian case. The second being diagrams that
contain vertex functions of the solvable theory. The Feynman rules
for the former directly follow from the linked cluster theorem. The
rules for the latter graphs follow from a set of problem-independent
skeleton diagrams that we present subsequently. This set of rules
then allows the computation of corrections at arbitrary order in perturbation
theory in a non-iterative manner. We find, however, that the iterative
procedure itself proves highly efficient. Applied to the special case
of the Ising model, it affords only a few more calculations than the
method that is specific to the problem \cite{Vasiliev74,Vasiliev75}.
In this particular case, the iterative procedure is closely related
to the known high temperature expansion \cite{Georges91}.

The remainder of the paper is organized as follows: In \prettyref{sub:Definition-field-theory}
we introduce the notation and the minimal extension of the diagrammatic
language required. \prettyref{sub:Perturbative-setting} sets up the
general perturbative problem and shows why the expansion of the partition
function and the generating functional of connected Green's functions
straight forwardly extends to the non-Gaussian setting. These preparations
are required to derive in \prettyref{sub:Recursion-for-Gamma} our
main result, a recursive algebraic equation that determines $\Gamma$.
\prettyref{sub:Graphical-rules-for-Gamma} interprets the algebraic
expressions in diagrammatic language, leading to a set of Feynman
rules to calculate all graphs contributing to $\Gamma$ at arbitrary
given order in a non-iterative manner. In \prettyref{sub:Illustrative-example}
we exemplify the algorithm for the simplest non-Gaussian extension
of a $\varphi^{3}$ theory. In \prettyref{sub:Application-to-Ising}
we demonstrate its application to the Ising model (which turns into
the Sherrington-Kirkpatrick spin glass model for Gaussian random couplings
\cite{Kirkpatrick1978}) to derive the Thouless-Anderson-Palmer mean-field
theory \cite{Thouless77_593} and its higher order corrections \cite{Nakanishi97_8085,Tanaka98_2302,Georges91,Vasiliev74,Vasiliev75}
in a purely diagrammatic manner.

\section{Results}

We here chose a notation that should be transparent with regard to
the nature of the problem (following to some extent \cite{Kleinert09}):
our results easily transfer to classical statistical mechanics, quantum
mechanics, quantum statistics, or quantum field theory. For clarity,
we here stick to the language of classical statistical field theory,
in particular to bosonic fields. In the following \prettyref{sub:Definition-field-theory}
we set up the language and define elementary quantities. 

\subsection{Definition of a field theory\label{sub:Definition-field-theory}}

We assume that the physical system has degrees of freedom $x\in\E$.
We may think of $x$ as a scalar, a vector (as in the example in \prettyref{sub:Application-to-Ising})
or a (possibly multi-component) field. The domain $\E$ must be chosen
accordingly. The system is described by the action $S$
\begin{eqnarray*}
S:\E & \mapsto & \R.
\end{eqnarray*}
with the property that a particular configuration of $x$ appears
with probability $p(x)\propto\exp(S(x))$. The partition function
$\Z$ is then given as
\begin{eqnarray}
\Z(j) & := & \int_{x}\,\exp(S(x)+j^{\T}x),\qquad Z(j):=\Z(j)/\Z(0),\label{eq:def_partition}
\end{eqnarray}
where we denote as $j^{\T}x$ the inner product on the space $\E$.
The symbol $\int_{x}$ stands for the sum over all configurations
of $x\in\E$, which technically may be a sum, an integral, or a path-integral,
depending on the space $\E$. We call $j$ the source field. The properly
normalized moment-generating function is denoted as $Z(j)$. Its logarithm
is the cumulant generating function
\begin{eqnarray}
W(j) & := & \ln\,Z(j),\qquad\llangle x^{n}\rrangle(j):=W^{(n)}(j),\label{eq:def_W}
\end{eqnarray}
also called generating functional of the connected Green's functions
in field theory; in statistical mechanics it is related to the Helmholtz
free energy $F(j)=\ln\Z(j)$ by $W(j)=F(j)-F(0)$. We denote the $n$-th
derivative of a function $f$ by $f^{(n)}$ and the $n$-th cumulant
as $\llangle x^{n}\rrangle.$

To define the effective action $\Gamma(x^{\ast})$, we eliminate the
dependence on the source field $j$ in favor of the mean value $\langle x\rangle$
\begin{eqnarray}
x^{\ast}(j) & : & =\langle x\rangle=\llangle x\rrangle=W^{(1)}(j),\label{eq:mean_of_j}
\end{eqnarray}
by a Legendre-Fenchel transform

\begin{eqnarray}
\Gamma(x^{\ast}) & : & =\sup_{j}\,j^{\T}x^{\ast}-W(j).\label{eq:def_gamma}
\end{eqnarray}
We provide the derivation of \prettyref{eq:def_gamma} in \prettyref{app:Definition-effective-action},
because we need an intermediate step as the starting point of the
perturbative expansion. To this end we employ the reciprocity property
of Legendre transforms
\begin{eqnarray}
j(x^{\ast}) & = & \Gamma^{(1)}(x^{\ast}),\label{eq:equn_of_state}
\end{eqnarray}
which, as derived in the appendix (with \prettyref{eq:pre_Legendre}),
leads to the integral-differential equation \cite[ Sec. 3.23.6]{Kleinert09}
\begin{eqnarray}
\exp\left(-\Gamma(x^{\ast})\right) & = & \Z(0)^{-1}\int_{\delta x}\,\exp\left(S(x^{\ast}+\delta x)+\Gamma^{(1)\T}(x^{\ast})\,\delta x\right).\label{eq:int_diff_gamma}
\end{eqnarray}
The latter expression is the starting point for the perturbative expansion
of $\Gamma$. It is also the natural starting point of a loopwise
expansion. The derivative operators that lead to $W^{(1)}$ and $\Gamma^{(1)}$
are given according to the space $\E$ as either partial derivatives
or, in the case of fields, functional derivatives.

\subsection{Perturbative problems\label{sub:Perturbative-setting}}

The equation of state \prettyref{eq:equn_of_state} provides us with
a self-consistency equation for the mean value $x^{\ast}$. The general
strategy is therefore to obtain an approximation of $\Gamma$ that
includes fluctuation corrections and then use the equation of state
to get an approximation for the mean value including these very corrections.
We will here derive a perturbative procedure to calculate approximations
of $\Gamma$ and will find the graphical rules for doing so. To solve
a problem perturbatively we decompose the action as 
\begin{eqnarray}
S(x) & = & S_{0}(x)+\epsilon V(x)\label{eq:def_S_pert}
\end{eqnarray}
with a part $S_{0}$ that can be solved exactly, that is we know its
corresponding cumulant generating function $W_{0}(j)$, and the remaining
terms collected in $\epsilon V(x)$. For the special case of $S_{0}$
being quadratic in the fields $x$, this leads to the well known result
that corrections to $\Gamma$ are composed of one-line irreducible
graphs only; our algorithm includes but extends this case to non-quadratic
$S_{0}$.

\begin{figure}
\begin{fmffile}{test7}
\fmfset{thin}{0.75pt}
\fmfset{decor_size}{4mm}
	\begin{eqnarray*}
    {\bf a)} \quad
	\epsilon V =
	\parbox{30mm}{
		\begin{fmfgraph*}(40,40)
			\fmfsurroundn{i}{3}
			\fmf{plain}{i1,v1,i2}
			\fmf{plain}{i3,v1}
		\end{fmfgraph*}
	}\\
	{\bf b)} \quad
    W_{0}=
	\parbox{15mm}{
		\begin{fmfgraph}(30,30)
			\fmfsurroundn{i}{2}
			\fmf{phantom}{i1,v,i2}
			\fmfv{d.s=circle, d.filled=empty}{v}
		\end{fmfgraph}
	} \mkern-18mu = \; \quad
	\parbox{10mm}{
		\begin{fmfgraph*}(30,30)
			\fmfsurroundn{i}{2}	
			\fmf{plain}{i2,v}
			\fmf{phantom}{v,i1}
			\fmfv{d.s=circle, d.filled=empty}{v}
			\fmflabel{j}{i2}
		\end{fmfgraph*}		
	} + {1 \over 2} \; \quad
	\parbox{15mm}{
		\begin{fmfgraph*}(30,30)
			\fmfsurroundn{i}{2}	
			\fmf{plain}{i2,v}
			\fmf{plain}{v,i1}
			\fmfv{d.s=circle, d.filled=empty}{v}
			\fmflabel{j}{i2}
			\fmflabel{j}{i1}
		\end{fmfgraph*}		
	} + {1 \over 3!} \quad
	\parbox{15mm}{
		\begin{fmfgraph*}(30,30)
			\fmfsurroundn{i}{3}	
			\fmf{plain}{i2,v,i3}
			\fmf{plain}{v,i1}
			\fmfv{d.s=circle, d.filled=empty}{v}
			\fmflabel{j}{i1}
			\fmflabel{j}{i2}
			\fmflabel{j}{i3}
		\end{fmfgraph*}		
	}
	\end{eqnarray*}
\end{fmffile}

\caption{\textbf{Expansion around a theory with first three non-vanishing cumulants.
a)} Bare vertex, here corresponding to a term $\propto x^{3}$. \textbf{b)}
Cumulants of the solvable part $W_{0}$ of the theory: the first three
cumulants are non-vanishing, corresponding to circles with one, two,
and three legs, respectively.\textbf{\label{fig:Expansion-example}}}
\end{figure}
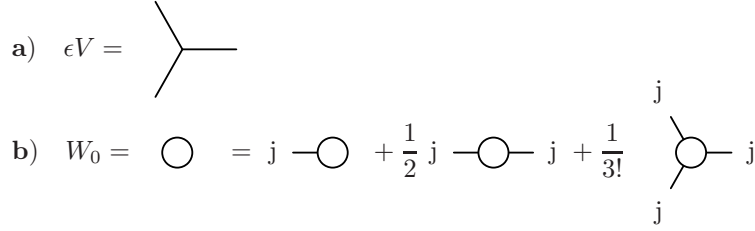

An example of this situation is depicted in \prettyref{fig:Expansion-example},
where the solvable theory $W_{0}$ already contains cumulants of order
one to three. We use the common notation for the interaction or bare
vertices, which we will call just ``vertices'' in the following:
the $n$-th power of the field $x$ is denoted as a vertex with $n$
emerging lines. The connecting elements in the graphs are the cumulants
contained in $W_{0}$. These are symbolized by circles, where the
number of legs corresponds to the order of the cumulant (following
e.g. \cite{Amit84,NegeleOrland98}). In ordinary Feynman diagrams,
$W_{0}$ would only have second order cumulants, which are typically
denoted by straight lines connecting the vertices.

The proof for the diagrammatic expansion closely follows the proof
of the linked cluster theorem (sometimes also called first Mayer theorem
\cite{Mayer77}), the connectedness of contributions to the cumulant
generating function $W$. We therefore briefly recapitulate the most
important steps of the latter proof (adapted from \cite[ Sec. 6.1.1]{ZinnJustin96})
here and provide all details in \prettyref{app:Perturbative-cumulant-expansion}
for completeness and consistency of notation.

The perturbative corrections $W_{V}(j)=W(j)-W_{0}(j)$ to the cumulant
generating function obey the operator equation (following from \prettyref{eq:perturbation_general})
\begin{eqnarray}
 &  & \exp(W_{V}(j))\label{eq:W_V_operator_eq}\\
 & = & \exp\left(-W_{0}(j)\right)\,\exp\left(\epsilon V(\partial_{j})\right)\,\exp\left(W_{0}(j)\right)\,\frac{\Z_{0}(0)}{\Z(0)},\nonumber 
\end{eqnarray}
where the latter constant factor is immaterial for the determination
of cumulants. Expanding in $\epsilon$, the function $W$ can be
constructed iteratively, where each step
\begin{eqnarray}
 &  & W_{l+1}(j)-W_{l}(j)\label{eq:expansion_W_iteration_main}\\
 & = & \frac{\epsilon}{L}\,\left(\exp\left(-W_{l}(j)\right)\,V(\partial_{j})\,\exp\left(W_{l}(j)\right)\right)\nonumber \\
 &  & +\mathcal{O}\big(\epsilon^{2}/L^{2}\big)\nonumber 
\end{eqnarray}
produces additional diagrams. The proof of connectedness then proceeds
by induction under the assumption that all graphs contained in $W_{l}$
are connected. In the following we will call a ``component'' a
term that appears in $W_{V}$ that is composed of a product of vertices,
the Taylor coefficients $V$, and of cumulants $W_{0}^{(n)}$, the
derivatives of the cumulant generating function of the unperturbed
theory. Representing $V$ in its Taylor series, every vertex of $V$
ties together components already contained in $W_{l}$. The cumulant
generating function is then given as
\begin{eqnarray*}
W & = & \lim_{L\to\infty}\,W_{L}.
\end{eqnarray*}
Combinatorics shows that only those graphs survive in the limit that
pick up each of their $k$ vertices in a different step in \prettyref{eq:expansion_W_iteration_main},
from which the factor $\left(\frac{\epsilon}{L}\right)^{k}\left(\begin{array}{c}
L\\
k
\end{array}\right)\;\stackrel{L\to\infty}{\to}\;\frac{\epsilon^{k}}{k!}$ arises (for details see \prettyref{app:Perturbative-cumulant-expansion}).

So we see that $W_{l}$ at any intermediate step $l$ may contain
connected components with arbitrary numbers of external legs. By the
inductive nature of the proof it is therefore clear that the linked
cluster theorem holds independently of the number of connected components
in $W_{0}$; that is to say it is generally true, even if already
$W_{0}$ contains connected components with arbitrary numbers of legs;
if it is the cumulant generating function of a non-Gaussian theory.

\subsection{Recursion for the effective action\label{sub:Recursion-for-Gamma}}

We now proceed to derive our main result, the diagrammatic expansion
of the effective action $\Gamma$. We here explain the main steps
and provide all details in \prettyref{app:operator_equation_Gamma_V}
and \prettyref{app:Recusion-equation-for-Gamma_V}.

To lowest order in perturbation theory, setting $\epsilon=0$ in \prettyref{eq:def_S_pert},
we get $W(j)=W_{0}(j)$; the leading order term in $\Gamma$ is the
corresponding Legendre transform
\begin{eqnarray}
\Gamma_{0}(x^{\ast}) & = & \sup_{j}\,j^{\T}x^{\ast}-W_{0}(j).\label{eq:Gamma0_pert-1}
\end{eqnarray}
We first need to derive a recursive equation to obtain approximations
of the form
\begin{eqnarray}
\Gamma(x^{\ast}) & = & :\Gamma_{0}(x^{\ast})+\Gamma_{V}(x^{\ast}),\label{eq:Gamma_pert_decomposition}
\end{eqnarray}
where we defined $\Gamma_{V}(x^{\ast})$ to contain all correction
terms that have at least one interaction vertex due to the interaction
potential $V$ to some order $\epsilon^{k}$ in perturbation theory.
We will use the iteration in a second step to proof Feynman rules
for the diagrams.

It is a priori not clear that the decomposition \prettyref{eq:Gamma_pert_decomposition}
is useful. We show in \prettyref{app:operator_equation_Gamma_V}
that this indeed so and, moreover, that it leads to the recursive
operator equation for $\Gamma_{V}$
\begin{eqnarray}
 &  & \exp(-\Gamma_{V}(x^{\ast}))\label{eq:Gamma_V_recursion}\\
 & = & \exp(-W_{0}(j))\,\exp\big(\epsilon V(\partial_{j})+\Gamma_{V}^{(1)\T}(x^{\ast})(\partial_{j}-x^{\ast})\big)\,\exp(W_{0}(j))\big|_{j=\Gamma_{0}^{(1)}(x^{\ast})},\nonumber 
\end{eqnarray}
which has a similar form as \prettyref{eq:W_V_operator_eq}, defining
the linked cluster expansion of $W$. The term $\Gamma_{V}^{(1)\T}(x^{\ast})(\partial_{j}-x^{\ast})$
can hence be seen as a monopole vertex. We want to solve the latter
equation iteratively order by order in the number of vertices $k$,
defining $\Gamma_{V,k}$. Analogous to the proof of the linked cluster
theorem, we arrive at a recursion by writing the exponential of the
differential operator in \prettyref{eq:Gamma_V_recursion} as a limit
and by expanding the logarithm to obtain the iteration
\begin{eqnarray}
 &  & g_{l+1}(j)-g_{l}(j)\label{eq:iteration_g}\\
 & = & \frac{\epsilon}{L}\,\exp(-W_{0}(j)-g_{l}(j))\,V(\partial_{j})\,\exp(W_{0}(j)+g_{l}(j))\label{eq:add_connected}\\
 & + & \frac{1}{L}\,\exp(-W_{0}(j)-g_{l}(j))\,\Gamma_{V}^{(1)}(x^{\ast})\left(\partial_{j}-x^{\ast}\right)\,\exp(W_{0}(j)+g_{l}(j))\label{eq:add_reducible}\\
 & + & \mathcal{O}(L^{-2}),\nonumber 
\end{eqnarray}
with initial condition
\begin{eqnarray}
g_{0} & = & 0.\label{eq:G_0}
\end{eqnarray}
We obtain the perturbation correction to $\Gamma$ \prettyref{eq:Gamma_pert_decomposition}
as the limit
\begin{eqnarray}
-\Gamma_{V}(x^{\ast}) & = & \lim_{L\to\infty}\,g_{L}(j)\Big|_{j=\Gamma_{0}^{(1)}(x^{\ast})}=:g(j)\Big|_{j=\Gamma_{0}^{(1)}(x^{\ast})}.\label{eq:Gamma_V_final}
\end{eqnarray}
To obtain the final result \prettyref{eq:Gamma_V_final}, we need
to express $j=\Gamma_{0}^{(1)}(x^{\ast})$ in $g_{l}(j)$. The latter
step is crucial to obtain $\Gamma_{V}(x^{\ast})$ as a function of
$x^{\ast}$. Because the mean value is given by $x^{\ast}=W_{0}^{(1)}(j_{0})$
for $j_{0}=\Gamma_{0}^{(1)}(x^{\ast})$ (following from \prettyref{eq:mean_of_j}
and \prettyref{eq:equn_of_state}), this step effectively expresses
all cumulants $\llangle x^{n}\rrangle_{0}\equiv W_{0}^{(n)}$ of
the unperturbed theory
\begin{eqnarray}
\llangle x^{n}\rrangle_{0}(x^{\ast}) & = & W_{0}^{(n)}(\Gamma_{0}^{(1)}(x^{\ast}))\label{eq:replacement_j}
\end{eqnarray}
in terms of the first cumulant $\llangle x\rrangle_{0}(x^{\ast})=x^{\ast}$,
where $x^{\ast}$ is the value of the first cumulant of the full theory.
The relation $j_{0}(x^{\ast})$ can practically be obtained either
by inverting $x^{\ast}=W_{0}^{(1)}(j_{0})$ or directly as $j_{0}=\Gamma_{0}^{(1)}(x^{\ast})$,
if $\Gamma_{0}(x^{\ast})$ is known explicitly. From the last point
follows that the differentiation in \prettyref{eq:add_reducible}
with $\Gamma_{V}^{(1)}(x^{\ast})\equiv\partial_{x^{\ast}}\Gamma_{V}(x^{\ast})=\partial_{x^{\ast}}(g\circ\Gamma_{0}^{(1)}(x^{\ast}))$
produces an inner derivative $\Gamma_{0}^{(2)}$ attached to a single
leg of a any unperturbed cumulant $W_{0}^{(n)}$ contained in $g_{L}$.
Defining the additional symbol
\begin{eqnarray*}
\Gamma_{0}^{(2)}(x^{\ast})=:\quad & \Diagram{g!p{0}g}
\end{eqnarray*}
allows us to write these contributions as
\begin{eqnarray}
\partial_{x^{\ast}}(g\circ\Gamma_{0}^{(1)}) & \equiv & (g^{(1)}\circ\Gamma_{0}^{(1)})\,\Gamma_{0}^{(2)}\Diagram{=\Diagram{!c{g}fg!p{0}g}
}
.\label{eq:subgraph_g_1-2}
\end{eqnarray}
This property will become important in the proof that the iteration
\prettyref{eq:iteration_g} generates only one-line irreducible graphs.
To see this, we have to generalize the notion of irreducibility known
from the Gaussian case, which we will do in the following section.

\subsection{Proof of a generalized irreducibility\label{sub:Graphical-rules-for-Gamma}}

The term ``one-line irredicibility'' in the Gaussian case refers
to the absence of diagrams that can be disconnected by cutting a single
second order bare propagator (a line in the original language of Feynman
diagrams). In the slightly generalized graphical notation introduced
in \prettyref{fig:Expansion-example}, these graphs would have the
form
\begin{eqnarray*}
 &  & \Diagram{!c{k^{\prime}}f!c{0}f!c{k^{\prime\prime}}\quad,}
\end{eqnarray*}
where two sub-graphs of $k$ and $k^{\prime}$ vertices, respectively,
are joined by a bare second order cumulant $\Feyn{fcf}$. Before
proceeding to the proof, we need to define irreducibility in a more
general sense than used in the Gaussian case. If a graph can be decomposed
into a pair of sub-graphs both containing vertices by disconnecting
the end point of a single vertex, we call this graph reducible. In
the Gaussian case, this definition is identical to one-line reducibility,
because all end points of vertices necessarily connect to a second
order propagator, a line. This is not necessarily the case if the
bare theory has higher order cumulants. We may have components of
graphs, such as
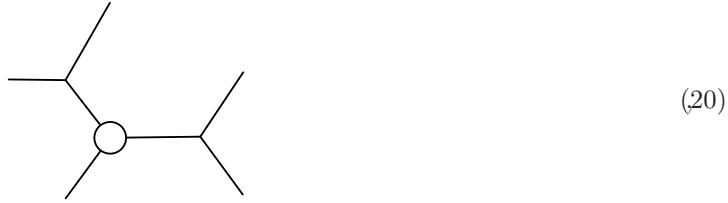
\begin{fmffile}{Example_three_point_cumu_three_point_int}	
	\begin{eqnarray}	
		&\parbox{100mm}{
			\begin{fmfgraph*}(150,75)
				\fmfpen{.75thin}
				\fmftop{ou1,og1,ou2,og2,ou3,og3,ou4,og4,ou5,og5}
				\fmfbottom{uu1,ug1,uu2,ug2,uu3,ug3,uu4,ug4,uu5,ug5}
				\fmf{phantom}{ou1,g1,G2,ug2}
				\fmf{phantom}{ou2,v1,g2,ug3}
				\fmf{plain}{v1,g2}
				\fmf{phantom}{ou3,G1,v2,ug4}
				\fmf{phantom}{ou4,g3,G3,ug5}
				\fmf{plain, tension=1.25}{v2,ug4}
				\fmf{phantom}{ou3,v1,G2,ug1}
				\fmf{plain, tension = 1.25}{ou3,v1}
				\fmf{phantom}{v1,G2,ug1}
				\fmf{phantom}{ou4,G1,g2,ug2}
				\fmf{plain}{g2,ug2}
				\fmf{phantom}{ou5,g3,v2,ug3}
				\fmf{phantom,tension=1.}{ou5,g3}
				\fmf{plain,tension=1.}{g3,v2}
				\fmf{plain}{g2,v2}
				\fmf{plain}{g1,v1}
				\fmfv{decor.shape=circle,decor.filled=empty, decor.size=12.thin}{g2} 
			\end{fmfgraph*}
		},&
	\end{eqnarray}
\end{fmffile}

where the three-point cumulant (could also be of higher order) connects
to two third (or higher) order interactions on either side. Disconnecting
a single leg, either to the left or to the right, decomposes the diagram
into two parts, each of which contains at least one interaction vertex.
We call such a diagram reducible and diagrams without this property
irreducible here. Note that a single leg of an interaction vertex
that ends on a first order cumulant does not make a diagram reducible;
both components need to contain at least one interaction.

\subsubsection{Cancellation of reducible diagrams\label{sec:Cancellation-of-reducible}}

We employ the following graphical notation: Since
\begin{eqnarray*}
g_{l}(j) & = & :\feyn{!c{g_{l}}}
\end{eqnarray*}
 depends on $j$ only indirectly by the $j$-dependence of the contained
bare cumulants, we denote the derivative by attaching one leg, which
is effectively attached to one of the cumulants of $W_{0}$ contained
in $g_{l}$ 
\begin{eqnarray*}
\Diagram{\vertexlabel^{j}f!c{g_{l}}}
 & := & \partial_{j}\:\Diagram{!c{g_{l}}}
:=\partial_{j}g_{l}(j).
\end{eqnarray*}

We first note that \prettyref{eq:iteration_g} generates two kinds
of contributions to $g_{l+1}$, corresponding to the lines \prettyref{eq:add_connected}
and \prettyref{eq:add_reducible}, respectively. The first line causes
contributions that come from the vertices of $\epsilon V(\partial_{j})$
alone. These are similar as in the linked cluster theorem \prettyref{eq:expansion_W_iteration_main}.
Determining the first order correction yields with $g_{0}=0$
\begin{eqnarray}
g_{1}(j) & = & \frac{\epsilon}{L}\,\exp(-W_{0}(j))\,V(\partial_{j})\,\exp(W_{0}(j))\label{eq:first_order_g1}\\
 & + & \mathcal{O}(L^{-2}),\nonumber 
\end{eqnarray}
which contains all graphs with a single vertex from $V$ and connections
formed by cumulants of $W_{0}$. These graphs are trivially irreducible,
because they only contain a single vertex.

The proof of the linked cluster theorem (see \prettyref{app:Perturbative-cumulant-expansion})
shows how the construction proceeds recursively: correspondingly the
$l+1$-st step \prettyref{eq:add_connected} generates all connected
graphs from components already contained in $W_{0}+g_{l}$. These
are tied together with a single additional vertex from $\epsilon V(x)$.
In each step, we only need to keep those graphs where the new vertex
in \prettyref{eq:add_connected} joins at most one component from
\textbf{$g_{l}$} to an arbitrary number of components of $W_{0}$,
hence we maximally increase the number of vertices in each component
by one. This is so, because comparing the combinatorial factors \prettyref{eq:factor_pick_up_one}
and \prettyref{eq:factor_pick_up_many}, contributions formed by adding
more than a single vertex (joining two or more components from $g_{l}$
by the new vertex) in a single step are suppressed with at least $L^{-1}$,
so they vanish in the limit \prettyref{eq:Gamma_V_final}.

The second term \prettyref{eq:add_reducible} is similar to \prettyref{eq:add_connected}
with three important differences:

\begin{itemize}
\item The differential operator appears in the form $\partial_{j}-x^{\ast}$.
As a consequence, when setting $j_{0}=\Gamma_{0}^{(1)}(x^{\ast})$
in the end in \prettyref{eq:Gamma_V_final}, all terms cancel where
$\partial_{j}$ acts directly on $W_{0}(j)$, because $W_{0}^{(1)}(j_{0})=x^{\ast}$;
non-vanishing contributions only arise if the $\partial_{j}$ acts
on a component contained in $g_{l}$. Since vertices and cumulants
can be composed to a final graph in arbitrary order, the diagrams
produced by $\partial_{j}-x^{\ast}$ acting on $g_{l}$ are the same
as those in which $\partial_{j}-x^{\ast}$ first acts on $W_{0}$
and in a subsequent step of the iteration another $\partial_{j}$
acts on the produced $W_{0}^{(1)}$. So to construct the set of all
diagrams it is sufficient to think of $\partial_{j}$ as acting on
$g_{l}$ alone; the reversed order of construction, where $\partial_{j}$
first acts on $W_{0}$ and in subsequent steps of the iteration the
remainder of the diagram is attached to the resulting $W_{0}^{(1)}$,
is contained in the combinatorics.
\item The single appearance of the differential operator $\partial_{j}$
acts like a monopole vertex: the term therefore attaches an entire
sub-diagram contained in $\Gamma_{V}^{(1)}$ by a single link to any
component contained in $g_{l}$.
\item These attached sub-diagrams from $\Gamma_{V}^{(1)}(x^{\ast})$ do
not depend on $j$; the $j$-dependence of all contained cumulants
is fixed to the value $j=\Gamma_{0}^{(1)}(x^{\ast})$, as seen from
\prettyref{eq:Gamma_V_recursion}. As a consequence, these sub-graphs
cannot form connections to vertices in subsequent steps of the iteration.
\end{itemize}
Considering that $\Gamma_{V}^{\left(1\right)}\left(x^{\ast}\right)$
is represented by \prettyref{eq:subgraph_g_1-2}, in step $l+1$ the
line \prettyref{eq:add_reducible} contributes graphs of the form
\begin{eqnarray}
g^{(1)}\,\Gamma_{0}^{(2)}\,g_{l}^{(1)} & =\Diagram{!c{g}fg!p{0}gf!c{g_{l}}}
 & \quad.\label{eq:reducible_general-1}
\end{eqnarray}
Since by their definition as a pair of Legendre transforms we have
\begin{eqnarray*}
1 & =\Gamma_{0}^{(2)}W_{0}^{(2)} & =\Diagram{g!p{0}gf!c{0}f}
\quad,
\end{eqnarray*}
we notice that the subtraction of the graphs \prettyref{eq:reducible_general-1}
may cancel certain connected graphs produced by the line \prettyref{eq:add_connected}.
In the case of a Gaussian solvable theory $W_{0}$ this cancellation
is the reason why only one-line irreducible contributions remain.
We here obtain the more general result, that these contributions cancel
all reducible components, according to the definition above. We will
prove this central result in the following.

To see the cancellation, we note that a reducible graph by our definition
has at least two components joined by a single leg of a vertex. Let
us first consider the case of a diagram consisting of exactly two
irreducible sub-diagrams joined by a single leg, as it is generated
by \prettyref{eq:add_reducible}. This leg may either belong to the
part $g^{(1)}$ or to $g_{l}^{(1)}$ in \prettyref{eq:reducible_general-1},
so either to the left or to the right sub-diagram. In both cases,
there is a second cumulant $W_{0}^{(2)}$ either left or right of
$\Gamma_{0}^{(2)}$. This is because if the two components are joined
by a single leg, this particular leg must have terminated on a $W_{0}^{(1)}$
prior to the formation of the compound graph; in either case this
term generates $W_{0}^{(1)}\stackrel{\partial_{j}}{\to}W_{0}^{(2)}$.

The second point to check is the combinatorial factor of graphs of
the form \prettyref{eq:reducible_general-1}. To construct a graph
of order $k$, where the left component has $k^{\prime}$ bare vertices
and the right has $k-k^{\prime}$, we can choose one of $L$ steps
within the iteration in which we may pick up the left term by \prettyref{eq:add_reducible}.
The remaining $k-k^{\prime}$ vertices are picked up by \prettyref{eq:add_connected},
which are $\left(\begin{array}{c}
L-1\\
k-k^{\prime}
\end{array}\right)$ possibilities to choose $k-k^{\prime}$ steps from $L-1$ available
ones. Every addition of a component to the graph comes with $L^{-1}$.
Any graph in $\Gamma_{V}$ with $k^{\prime}$ vertices is $\propto\frac{\epsilon^{k^{\prime}}}{k^{\prime}!}$,
so together we get 
\begin{eqnarray}
\frac{L}{L}\,\frac{\epsilon^{k^{\prime}}}{k^{\prime}!}\,\left(\frac{\epsilon}{L}\right)^{k-k^{\prime}}\left(\begin{array}{c}
L-1\\
k-k^{\prime}
\end{array}\right) & \stackrel{L\to\infty}{\to} & \frac{\epsilon^{k}}{k^{\prime}!(k-k^{\prime})!}.\label{eq:comb_reducible-1}
\end{eqnarray}

The symmetry factors $s_{1},s_{2}$ of the two sub-graphs generated
by \prettyref{eq:reducible_general-1} enter the symmetry factor $s=s_{1}\cdot s_{2}\cdot c$
of the composed graph as a product, where $c$ is the number of ways
in which the two sub-graphs may be joined. But the factor $s$, by
construction, excludes those symmetries that interchange vertices
between the two sub-graphs. Assuming, without loss of generality,
a single sort of interaction vertex, there are $s^{\prime}=\left(\begin{array}{c}
k\\
k^{\prime}
\end{array}\right)=\frac{k!}{k^{\prime}!(k-k^{\prime})!}$ ways of choosing $k^{\prime}$ of the $k$ vertices to belong to
the left part of the diagram. Therefore the symmetry factor $s$ is
smaller by the factor $s^{\prime}$ than the symmetry factor of the
corresponding reducible diagram constructed by \prettyref{eq:add_connected}
alone, because the latter exploits all symmetries, including those
that mix vertices among the sub-graphs. Combining the defect $s^{\prime}$
with the combinatorial factor \prettyref{eq:comb_reducible-1} yields
$\frac{1}{k^{\prime}!(k-k^{\prime})!}/s^{\prime}=\frac{1}{k!}$, which
equals the combinatorial factor of the reducible graph under consideration. 

The generalization to the case of an arbitrary number of sub-diagrams
of which $M$ are irreducible and connected to the remainder of the
diagram by exactly one link, called ``leaves'', is straightforward:
we can always pick one of these sub-diagrams and replace it by its
corresponding contribution to $\Gamma^{\left(1\right)}\left(x^{\ast}\right)$.
Then, by the same arguments as before, we produce the same diagram
with the same prefactor, only with opposite sign. In summary we conclude
that all reducible graphs are canceled by \prettyref{eq:reducible_general-1}.
We therefore obtain the first part of our Feynman rules: All connected,
irreducible diagrams that are also contained in the perturbation expansion
of $W_{V}$ also contribute to $\Gamma_{V}$ with a negative sign
and the same combinatorial factor.

But there is a second sort of graphs produced by \prettyref{eq:reducible_general-1}
that does not exist in the Gaussian case: If the connection between
the two sub-components by $\Feyn{gpg}$ ends on a third or higher
order cumulant. These graphs cannot be produced by \prettyref{eq:add_connected},
so they remain with a minus sign. We show an example of such graphs
in \prettyref{fig:reducible-diagrams-generated}c). One might wonder
why this contribution does not cancel while making a ``reducible
impression'', if one interprets $\Gamma_{0}^{\left(2\right)}$ as
a kind of two-point interaction. The solution is that for diagrams
of this kind, indeed contributions of opposite signs and same absolute
values are generated, but the contribution with the one sign is generated
once and the contribution with the other sign twice, therefore the
total contribution does not vanish. We address this issue more thoroughly
in \prettyref{subsec:Taxonomy-of-reducible}.

Moreover, subsequent application of $\partial_{x^{\ast}}$ on such
components produces graphs that contain higher order derivatives of
$\Gamma_{0}^{(n)}$. By the arguments given in the proof for the generalized
irreducibility, we deduce that all diagrams cancel that have at least
one leaf from $W_{V}^{\left(1\right)}$ (that includes only unperturbed
cumulants and interactions, but no vertices $\Gamma^{\left(n\right)}\left(x^{\ast}\right)$)
and that is connected to the remainder of the diagram by a single
leg of an interaction (and not via a leg of $\Gamma^{\left(n\right)}\left(x^{\ast}\right)$,
see \prettyref{subsec:Taxonomy-of-reducible}). Reducible non-standard
diagrams that violate these prerequisites can occur. In the end of
the following subsection, we will also give an example for this case.
To enumerate all all non-standard diagrams and to find their Feynman
rules, we also derive a set of skeleton diagrams in the following
subsection that allow the computation of all contributions at any
given order including their combinatorial factor. 

\subsubsection{Perturbative diagrammatics derived from skeleton diagrams\label{sec:Perturbative-diagrammatics-deriv}}

To obtain a better understanding of the types of diagrams that contribute
to the perturbation expansion, in particular the non-standard diagrams
not contained in $W_{V}$, it is useful to derive a non-iterative
approach based on a set of skeleton diagrams. We use the term ``skeleton
diagram'' for a diagram containing dressed as opposed to bare cumulants
and vertices. We start with a Taylor expansion around a adroitly chosen
point $x_{1}$
\begin{eqnarray}
\Gamma(x^{\ast}) & = & \Gamma\left(x_{1}\right)+\Gamma^{\left(1\right)}\left(x_{1}\right)^{\T}\left(x^{\ast}-x_{1}\right)\label{eq:Taylor_Gamma_Skeleton}\\
 &  & +\sum_{n=2}\,\sum_{i_{1}\cdots i_{n}}\frac{\Gamma_{i_{1}\cdots i_{n}}^{(n)}(x_{1})}{n!}\,\prod_{l=1}^{n}(x^{\ast}-x_{1})_{i_{l}},\nonumber 
\end{eqnarray}
where we wrote the first two terms explicitly. We now choose a particular
point $x_{1}$ by making use of the involutive property of the Legendre
transform, which is given for all smooth and convex cumulant generating
functions, irrespective of the form of the underlying theory. Concretely,
we choose $x_{1}:=W^{(1)}(\Gamma_{0}^{(1)}(x^{\ast})),$ which is
equivalent to
\begin{equation}
\begin{array}{lcr}
j_{0} & = & \Gamma_{0}^{(1)}(x^{\ast})\\
j_{0} & = & \Gamma^{(1)}(x_{1})
\end{array}\Longleftrightarrow\begin{array}{lcr}
x^{\ast} & = & W_{0}^{(1)}(j_{0})\\
x_{1} & = & W^{(1)}(j_{0})
\end{array}.\label{eq:Def_x1_via_x_star}
\end{equation}
As an immediate consequence, we see that
\begin{equation}
x^{\ast}-x_{1}=W_{0}^{(1)}(j_{0})-W^{(1)}(j_{0})=-W_{V}^{(1)}(j_{0}).\label{eq:W_V_x}
\end{equation}
Using these relations and the definition of the Legendre transform,
the two terms in the first line of \prettyref{eq:Taylor_Gamma_Skeleton}
can be rewritten as
\begin{eqnarray}
 &  & j_{0}^{\T}x_{1}-W\left(j_{0}\right)+j_{0}^{\T}\,\left(x^{\ast}-x_{1}\right)\nonumber \\
 & = & j_{0}^{\T}x^{\ast}-W_{0}\left(j_{0}\right)-W_{V}\left(j_{0}\right)\nonumber \\
 & = & \Gamma_{0}\left(x^{\ast}\right)-W_{V}\left(j_{0}\right).\label{eq:Vertex_Taylor_x1}
\end{eqnarray}
We now see that the first two terms in \prettyref{eq:Taylor_Gamma_Skeleton},
besides $\Gamma_{0}(x^{\ast})$, also contain all diagrams from $W_{V}\left(j_{0}\right)$,
but with the opposite sign. We note that by their dependence on $x^{\ast}$
and $j_{0}$, respectively, these contributions can be calculated:
The relation \prettyref{eq:Def_x1_via_x_star} allows us to express
all bare cumulants $W_{0}^{(n)}(j_{0})$ that appear in $-W_{V}$
in terms of the first cumulant $x^{\ast}\equiv W_{0}(j_{0})$. 

The second line in \prettyref{eq:Taylor_Gamma_Skeleton} produces
additional diagrams that are of second order in the interaction or
higher; this is because the ``leaves'' \prettyref{eq:W_V_x} of
these tree diagrams are $-W_{V}$, which, by definition, contain at
least one interaction vertex. These terms include those diagrams that
cancel the reducible diagrams included in $W_{V}\left(j_{0}\right)$,
as proven above. In addition, these terms contain the non-standard
diagrams. We will describe in the following, how the exact form and
combinatorial factors of these diagrams can be obtained. 

To express the dressed vertex functions $\Gamma^{(n)}(x_{1})$ that
appear in the second line of \prettyref{eq:Taylor_Gamma_Skeleton},
we first derive a perturbative expansion for $\Gamma^{(2)}(x_{1})$,
which is basically Dyson's equation. We here include the derivation
to see that it holds beyond the Gaussian case. Decomposing $W$ into
the solvable part and its perturbative corrections, the reciprocity
relation 
\begin{eqnarray}
W^{(2)}(j_{0})\,\Gamma^{(2)}(W^{(1)}(j_{0})) & = & 1\label{eq:reciprocity-1}
\end{eqnarray}
takes the form
\begin{eqnarray*}
(W_{0}^{(2)}(j_{0})+W_{V}^{(2)}(j_{0}))\,\Gamma^{(2)}(\underbrace{W^{(1)}(j_{0})}_{\equiv x_{1}}) & = & 1.
\end{eqnarray*}
Multiplying from left with $\Gamma_{0}^{(2)}(W_{0}^{(1)}(j_{0}))=\Gamma_{0}^{(2)}(x^{\ast})$,
which is the inverse of $W_{0}^{(2)}(j_{0})$, we obtain by rearranging
\begin{eqnarray}
\Gamma^{(2)}(x_{1}) & = & \Gamma_{0}^{(2)}(x^{\ast})-\Gamma_{0}^{(2)}(x^{\ast})\,W_{V}^{(2)}(j_{0})\,\Gamma^{(2)}(x_{1}).\label{eq:Dyson_recursion}
\end{eqnarray}
The latter term contains at least one interaction vertex in $W_{V}$,
the former is of order zero. So we may iterate this equation to obtain
an inverted Dyson's equation for $\Gamma^{(2)}$ that reads
\begin{eqnarray}
\Gamma^{(2)}(x_{1}) & = & \Gamma_{0}^{(2)}(x^{\ast})-\Gamma_{0}^{(2)}(x^{\ast})\,W_{V}^{(2)}(j_{0})\,\Gamma_{0}^{(2)}(x^{\ast})\pm\ldots\label{eq:inversed_Dyson}\\
\nonumber \\
 & = & \Diagram{g!p{0}g-g!p{0}g!c{V}g!p{0}g}
\pm\ldots.\nonumber 
\end{eqnarray}
It is easily checked by insertion that \prettyref{eq:inversed_Dyson}
solves \prettyref{eq:Dyson_recursion}.

So we have expressed the second order vertex at $x_{1}$ by means
of quantities that are directly accessible or can be calculated perturbatively
in a straight-forward way, by their dependence on $x^{\ast}$ and
$j_{0}$, respectively. Higher order vertices are determined in the
standard way by differentiating the reciprocity relation \prettyref{eq:reciprocity-1}
multiple times with respect to $j_{0}$, and removing the legs $W^{(2)}(j_{0})$
that arise from inner derivatives by multiplication with the corresponding
inverse $\Gamma^{(2)}(x_{1})$. The result is the well-known decomposition
of vertex in terms of tree skeleton diagrams \cite{ZinnJustin96,NegeleOrland98}

\begin{eqnarray}
\Gamma_{123}^{(3)}(x_{1}) & = & -\sum_{\{i_{l}\}}W_{i_{1}i_{2}i_{3}}^{(3)}(j_{0})\,\Gamma_{i_{1}1}^{(2)}(x_{1})\,\Gamma_{i_{2}2}^{(2)}(x_{1})\,\Gamma_{i_{3}3}^{(2)}(x_{1})\label{eq:skeleton}\\
\Gamma_{1234}^{(4)}(x_{1}) & = & -\sum_{\{i_{l}\}}W_{i_{1}i_{2}i_{3}i_{4}}^{(4)}(j_{0})\,\Gamma_{i_{1}1}^{(2)}(x_{1})\,\Gamma_{i_{2}2}^{(2)}(x_{1})\,\Gamma_{i_{3}3}^{(2)}(x_{1})\,\Gamma_{i_{4}4}^{(2)}(x_{1})\nonumber \\
 &  & +\sum_{\{i_{l}\}}W_{i_{1}i_{2}i_{3}}^{(3)}\,W_{i_{4}i_{5}i_{6}}^{(3)}(j_{0})\,\Gamma_{i_{1}1}^{(3)}(x_{1})\,\Gamma_{i_{2}2}^{(2)}(x_{1})\,\Gamma_{i_{3}i_{4}}^{(2)}(x_{1})\,\Gamma_{i_{5}3}^{(2)}(x_{1})\,\Gamma_{i_{6}4}^{(2)}(x_{1})+2\,\text{perm.}\nonumber \\
 & \ldots & .\nonumber 
\end{eqnarray}
Diagrammatically, the first of these relations reads

\begin{fmffile}{treestructure_of_vertices_2}
\fmfset{thin}{0.75pt}
\fmfset{decor_size}{4mm}
	\begin{eqnarray*}
	\parbox{30mm}{
		\begin{fmfgraph*}(40,40)
			\fmfsurroundn{i}{3}
			\fmf{plain}{i1,v1,i2}
			\fmf{plain}{i3,v1,}
			\fmflabel{1}{i1}
			\fmflabel{2}{i2}
			\fmflabel{3}{i3}
			\fmfv{decor.shape=circle, d.filled=empty}{v1}
		\end{fmfgraph*}
		} =& \; - \; \parbox{30mm}{
		\begin{fmfgraph*}(100,100)
			\fmfsurroundn{i}{3}
			\fmf{wiggly}{v1,v3}
			\fmf{wiggly}{v1,v5}
			\fmf{wiggly}{v1,v7}
			\fmf{plain}{i1,v2,v3}
			\fmf{plain}{i2,v4,v5}
			\fmf{plain}{i3,v6,v7}
			\fmflabel{1}{i1}
			\fmflabel{2}{i2}
			\fmflabel{3}{i3}
			\fmfv{label=$i_1$, l.a=90}{v3}
			\fmfv{label=$i_2$, l.a=-170}{v5}
			\fmfv{label=$i_3$, l.a=-10}{v7}
			\fmfv{decor.shape=circle, d.filled=shaded}{v1}
			\fmfv{decor.shape=circle, d.filled=empty}{v2}
			\fmfv{decor.shape=circle, d.filled=empty}{v4}
			\fmfv{decor.shape=circle, d.filled=empty}{v6}
		\end{fmfgraph*}
		}
	\end{eqnarray*}
\end{fmffile}We can hence express vertices of arbitrary order at the point $x_{1}$
by cumulants at the point $j_{0}$ (which is given by $x^{\ast}$
by means of \prettyref{eq:Def_x1_via_x_star}) and $\Gamma^{\left(2\right)}\left(x_{1}\right)$,
which is again given by the same cumulants and $\Gamma_{0}^{(2)}(x^{\ast})$
by Dyson's equation \prettyref{eq:inversed_Dyson}. The important
point to note here is that the set of rules to translate skeleton
diagrams into their perturbative expansion is problem independent;
it just results from the properties of the Legendre transform and
the decomposition into solvable part and perturbation.

Because both factors in every term of the sum in the second line of
\prettyref{eq:Taylor_Gamma_Skeleton} depend on the interaction $V$,
the order of each such term is the sum of the orders of its factors.
We derive two additional rules from this view. First, since also $W_{0}(j_{0})$
and $\Gamma_{0}(x^{\ast})$ form a pair of Legendre transforms, their
reciprocity relation $W_{0}^{(2)}(j_{0})\,\Gamma_{0}^{(2)}(x^{\ast})=1$
gives rise to a relation corresponding to \prettyref{eq:skeleton},
but with $\Gamma^{(n)}(x_{1})$ replaced by $\Gamma_{0}^{(n)}(x^{\ast})$
and $W^{(n)}(j_{0})$ by $W_{0}^{(n)}(j_{0})$. This means that the
lowest order terms for the $n$-th derivative in \prettyref{eq:skeleton}
resum to $\Gamma_{0}^{(n)}(x^{\ast})$. The final set of skeleton
diagrams derived from \prettyref{eq:Taylor_Gamma_Skeleton}, the main
result of this section therefore reads
\begin{eqnarray}
\Gamma(x^{\ast}) & = & \Gamma_{0}(x^{\ast})-W_{V}(j_{0})\label{eq:Gamma_skeleton_final}\\
 &  & +\sum_{n=2}\,\frac{\Gamma_{0}^{(n)}(x^{\ast})+\bar{\Gamma}^{(n)}(x_{1})}{n!}\,(-W_{V}^{(1)}(j_{0}))^{n},\nonumber 
\end{eqnarray}
where we defined $\bar{\Gamma}^{(n)}(x_{1}):=\Gamma^{(n)}(x_{1})-\Gamma_{0}^{(n)}(x^{\ast})$,
which contains all diagrams from \prettyref{eq:inversed_Dyson} and
\prettyref{eq:skeleton} that have at least one interaction vertex,
so which are of order $\mathcal{O}(\epsilon)$ or higher.

The second rule follows from $W_{V}^{(1)}(j_{0})$ being of first
order or higher, so that at the $n$-th order in the interaction,
we have to consider at most the $n$-th term in the sum in \prettyref{eq:Gamma_skeleton_final}.
In this last term we must in addition drop the $\bar{\Gamma}$-term.
The only term to consider at second order, for example, is
\[
\frac{1}{2!}\,W_{V}^{(1)}(j_{0})^{\T}\,\Gamma^{(2)}(x_{1})\,W_{V}^{(1)}(j_{0})=\frac{1}{2!}\,W_{V}^{(1)\T}(j_{0})\Gamma_{0}^{(2)}(x^{\ast})W_{V}^{(1)}(j_{0})+\mathcal{O}\left(\epsilon^{3}\right).
\]
This term cancels the contributions of the diagrams b) and c) in \prettyref{fig:Diagrams-connected-second-order}.
However, that all reducible diagrams cancel can be seen more systematically
by the constructive arguments given in \prettyref{sec:Cancellation-of-reducible}.

But eq. \prettyref{eq:Vertex_Taylor_x1} is useful to get a handle
on the non-standard diagrams. If we want to know all such diagrams
of a certain order $k$ in the interaction, we first assign the number
of interactions to be contained in each factor in any term $\Gamma^{(n)}(x_{1})/n!\,\left(-W_{V}^{(1)}(j_{0})\right)^{n}$,
$2\leq n\leq k$ so that they sum up to $k$. Assigning $0$ interaction
vertices to $\Gamma^{(n)}$ reduces it to $\Gamma_{0}^{(n)}(x^{\ast})$,
as said above. For non-zero numbers of vertices in $\bar{\Gamma}^{(n)}$,
we decompose the latter via \prettyref{eq:inversed_Dyson} and \prettyref{eq:skeleton}
into an expression only containing full cumulants and $\Gamma_{0}^{\left(2\right)}\left(x^{\ast}\right)$
and distribute again the number of interactions assigned to $\bar{\Gamma}^{(n)}$
among these cumulants. Each full cumulant is then broken down into
all its perturbative expansion, for which the standard Feynman rules
hold according to the linked cluster theorem (see \prettyref{app:Perturbative-cumulant-expansion}).
Throughout this expansion, we skip all diagrams that are also contained
in $-W_{V}$, because these are just the reducible ones that we know
to cancel each other.

Another insight we gain by \prettyref{eq:Gamma_skeleton_final} in
combination with \prettyref{eq:inversed_Dyson} is that non-standard
diagrams are not necessarily irreducible - in fact, they can even
be reducible in the sense that they fall into two parts both containing
interactions if we cut a second-order cumulant. We show an example
for this phenomenon in a toy model at the end of \prettyref{sub:Illustrative-example}.

\subsubsection{Summary of Feynman rules for the non-Gaussian case}

We now summarize the algorithmic rules derived from the above observations
to obtain $\Gamma$:
\begin{enumerate}
\item \label{enu:Calculate-Gamma0}Calculate $\Gamma_{0}(x^{\ast})=\sup_{j}\,j^{\T}x^{\ast}-W_{0}(j)$
explicitly by finding $j_{0}$ that extremizes the right hand side.
At this order $g_{0}=0$.
\item \label{enu:recursion}At order $k$ in the perturbation expansion:
\begin{enumerate}
\item \label{enu:connected}add all irreducible graphs in the sense of the
definition above that have $k$ vertices;
\item \label{enu:add-all-irreducible}add all graphs containing derivatives
$\Gamma_{0}^{(n)}$ as connecting elements that cannot be reduced
to the form of a graph contained in the expansion of $W_{V}(j_{0})$;
the graphs left out are the counterparts of the reducible ones in
$W_{V}(j_{0})$. The topology and combinatorial factors of these non-standard
contributions are generated iteratively by \prettyref{eq:iteration_g}
from the previous order in perturbation theory; this iteration, by
construction, only produces diagrams, where at least two legs of each
$\Gamma_{0}^{(n)}$ connect to a third or higher order cumulant. We
can also directly leave out diagrams, in which a subdiagram contained
in $W_{V}$ is connected to the remainder of the diagram by a single
leg of an interaction vertex. Alternatively, the additional diagrams
can be constructed directly for any order $k$ by instantiating the
skeleton diagrams generated by \prettyref{eq:Gamma_skeleton_final}
to the desired order, as explained in \prettyref{sec:Perturbative-diagrammatics-deriv}.
Note that here diagrams may appear in which $\Gamma^{(n>2)}$ couples
to less than two third or higher order cumulants. After decomposing
all $\Gamma^{(n>2)}$ into $\Gamma_{0}^{\left(2\right)}$ and diagrams
from $W^{\left(n\right)}$ by \prettyref{eq:skeleton} and \prettyref{eq:inversed_Dyson},
we can discard diagrams, in which at least one subdiagram from $W_{V}$
forms a leave that is connected by a single link of an interaction
vertex; these cancel by the same argument as given in the iterative
approach. 
\end{enumerate}
\item \label{enu:assign-the-factor}assign the factor $\frac{\epsilon^{k}}{r_{1}!\cdots r_{l+1}!}$
to each diagram with $r_{i}$-fold repeated occurrence of vertex $i$;
assign the combinatorial factor that arises from the possibilities
of joining the connecting elements as usual in Feynman diagrams (see
examples below and e.g. \cite{Amit84});
\item \label{enu:j0_as_x_star}express the $j$-dependence of the $n$-th
cumulant $\llangle x^{n}\rrangle(x^{\ast})$ in all terms by the first
cumulant $x^{\ast}=\llangle x\rrangle=W_{0}^{(1)}(j_{0})$; this can
be done, for example, by inverting the last equation or directly by
using $j_{0}=\Gamma_{0}^{(1)}(x^{\ast})$; express the occurrence
of $\Gamma_{0}^{(2)}$ by its explicit expression.
\end{enumerate}

This set of Feynman rules constitutes the central result of our work.
Note that the rules include the well-known Gaussian case, because
irreducibility in the here-defined sense is identical to one-line-irreduciblity
in the expansion around a Gaussian theory: every leg of an interaction
vertex necessarily connects to a line there. Non-standard diagrams
hence cannot appear in this case.

The Feynman rules hold for any order $k$. So one may compute corrections
at order $k$ directly without having computed any lower order. The
combinatorial factor is fixed unambiguously, too. To get an intuitive
understanding of the cancellation, it is still useful to illustrate
the recursive construction of graphs in \prettyref{sub:Illustrative-example}
in the application to a minimal, non-Gaussian setting. In \prettyref{sub:Application-to-Ising}
we demonstrate the application to systems of Ising spins, recovering
the TAP approximation \cite{Thouless77_593}, the high temperature
expansion \cite{Georges91}, and the Plefka expansion \cite{Plefka82_1971}.
It turns out that the recursive algorithmic procedure leads to a dramatic
decrease of required computations.

\subsection{Illustrative example for the graphical rules\label{sub:Illustrative-example}}

As a first example let us consider a zero-dimensional field theory,
that is to say, a probability distribution of a scalar variable. By
assumption, $S_{0}$ constitutes the solvable theory, so that $W_{0}(j)$
can be determined exactly by \prettyref{eq:def_W}. We here illustrate
the method for a solvable theory that has non-vanishing cumulants
of orders one, two, and three, hence $W_{0}$ is a polynomial of order
three

\begin{eqnarray*}
W_{0}(j) & = & \feyn{!c{0}}=\sum_{n=1}^{3}\,\llangle x^{n}\rrangle_{0}\,\frac{j^{n}}{n!},
\end{eqnarray*}
where the cumulants $\llangle x^{n}\rrangle_{0}$ with $n\in\{1,2,3\}$
appear as Taylor coefficients. Its graphical representation is shown
in \prettyref{fig:Expansion-example}b. As perturbation we assume
a three point vertex $\epsilon V(x)=\epsilon\,x^{3}$, shown in \prettyref{fig:Expansion-example}a.

According to step (\prettyref{enu:Calculate-Gamma0}) we determine
$\Gamma_{0}(x^{\ast})$ by \prettyref{eq:def_gamma}, which amounts
to the calculation of the derivative $\partial_{j}W_{0}(j_{0})-x^{\ast}=0$
which determines $j_{0}(x^{\ast})$.

We now calculate the diagrams recursively according to step (\prettyref{enu:recursion}).
At order $l=0$ we have $g_{0}=0$. At first order $l=1$, we therefore
get from \prettyref{eq:first_order_g1} the graphs shown in \prettyref{fig:Diagrams-first-order}.

\begin{figure}[H]
\begin{fmffile}{test8}
\fmfset{thin}{0.75pt}
\fmfset{decor_size}{4mm}
	{\bf a)} \quad $1 \cdot$ \quad
	 \parbox{15mm}{
		\begin{fmfgraph*}(30,30)
			\fmfsurroundn{i}{3}	
			\fmf{plain}{i2,v,i3}
			\fmf{plain}{v,i1}
			\fmfv{d.s=circle, d.filled=empty}{i1,i2,i3}
		\end{fmfgraph*}		
	} \qquad
	{\bf b)} \quad $3 \cdot$ \quad
	\parbox{20mm}{
		\begin{fmfgraph*}(40,30)
			\fmfsurroundn{i}{2}	
			\fmf{plain, tension=1.5}{i2,v}
			\fmf{plain, left=.7, tension=0.5}{v,i1,v}
			\fmfv{d.s=circle, d.filled=empty}{i1,i2}
		\end{fmfgraph*}		
	} \qquad
    {\bf c)} \quad $1 \cdot$ \quad
	\parbox{15mm}{
		\begin{fmfgraph*}(30,30)
			\fmfsurroundn{i}{2}	
			\fmf{plain}{i2,i1}
			\fmf{plain, left=.7, tension=0.5}{i2,i1,i2}
			\fmf{plain, tension=0.2}{i1,i2}
			\fmfv{d.s=circle, d.filled=empty}{i2}
		\end{fmfgraph*}		
	}
\end{fmffile}

\caption{All diagrams contributing to $g_{1}$ at first order $l=1$ generated
by \prettyref{eq:first_order_g1} for the theory of \prettyref{fig:Expansion-example}.\label{fig:Diagrams-first-order}}
\end{figure}

Each graph has a single vertex, thus we get a factor $\frac{\epsilon}{1!}$.
The combinatorial factors for each graph are stated explicitly above.
So the diagrams in $\Gamma_{V}$ at first order are identical to all
connected diagrams with a single vertex; in this regard, the expansion
is identical to the well known Gaussian case, except that the diagram
\prettyref{fig:Diagrams-first-order} c) would not appear in the absence
of third order cumulants of the solvable theory.

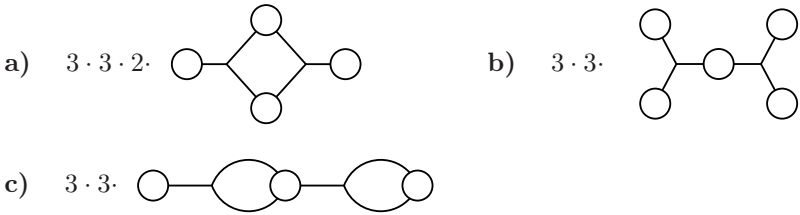
\begin{figure}[H]
\begin{fmffile}{test9a}
\fmfset{thin}{0.75pt}
\fmfset{decor_size}{4mm}
	{\bf a)} \quad $3 \cdot 3 \cdot 2 \cdot$ \quad
	\parbox{30mm}{
		\begin{fmfgraph*}(60,60)
			\fmfsurroundn{i}{4}
			\fmf{plain, tension=2}{i3,v1}	
			\fmf{plain}{v1,v2,v3,v4,v1}
			\fmf{plain, tension=2}{i1,v3}
			\fmf{phantom, tension=2.5}{i2,v2}
			\fmf{phantom, tension=2.5}{i4,v4}
			\fmfv{d.s=circle, d.filled=empty}{i3,i1,v2,v4}
		\end{fmfgraph*}		
	} \qquad
    {\bf b)} \quad $3 \cdot 3 \cdot$ \quad
	\parbox{30mm}{
		\begin{fmfgraph*}(60,30)
			\fmfleft{i1,i2}
			\fmfright{o1,o2}
			\fmf{plain, tension=1}{i1,v1,i2}	
			\fmf{plain}{v1,v2,v3}
			\fmf{plain, tension=1}{o1,v3,o2}
			\fmfv{d.s=circle, d.filled=empty}{i1,i2,v2,o1,o2}
		\end{fmfgraph*}		
	}\\
    {\bf c)} \quad $3 \cdot 3 \cdot$ \quad
	\parbox{40mm}{
		\begin{fmfgraph*}(100,30)
			\fmfleft{i1}
			\fmfright{o1}
			\fmf{plain, tension=2.5}{i1,v1}	
			\fmf{plain, left=.7}{v1,v2,v1}
			\fmf{plain, tension=2.5}{v2,v3}
			\fmf{plain, left=.7}{v3,o1,v3}
			\fmfv{d.s=circle, d.filled=empty}{i1,v2,o1}
		\end{fmfgraph*}		
	}
\end{fmffile}

\caption{Some diagrams generated by \prettyref{eq:add_connected} contributing
to $g_{2}-g_{1}$ at order $l=2$. Diagrams b) and c) are reducible,
so they will drop out.\label{fig:Diagrams-connected-second-order}}
\end{figure}

At order $l=2$, the graphs in $W_{0}\cup g_{1}$ contain those that
have been produced in the previous step. The step \prettyref{eq:add_connected}
therefore produces additional diagrams out of these components, some
of which are shown in \prettyref{fig:Diagrams-connected-second-order}.

The diagrams a) and b) in \prettyref{fig:Diagrams-connected-second-order}
are composed of the diagram a) in \prettyref{fig:Diagrams-first-order}
contained in $g_{1}$ and one additional vertex. The diagram c) in
\prettyref{fig:Diagrams-connected-second-order} is composed of diagram
b) in \prettyref{fig:Diagrams-first-order}; its combinatorial factor
$3\cdot3$ is the combined factor of the first order diagram and the
factor $3$ due to three possibilities to select a leg of the vertex.
We here skipped further diagrams; in particular all diagrams, that
are generated by the first order diagram c) in \prettyref{fig:Diagrams-first-order}.

The line \prettyref{eq:add_reducible} produces additional diagrams
with negative sign. We here indicate the simplification $\Feyn{gpgfcf}=1$
by an overbrace. Some diagrams are shown in \prettyref{fig:reducible-diagrams-generated}.

\begin{figure}[H]
\begin{fmffile}{test10}
\fmfset{thin}{0.75pt}
\fmfset{decor_size}{4mm}
	{\bf a)} \quad $- 3 \cdot 3 \cdot$
	\parbox{30mm}{
		\begin{fmfgraph*}(100,30)
			\fmfleft{i1,i2}
			\fmfright{o1,o2}
			\fmf{plain, tension=1}{i1,v1,i2}	
			\fmf{plain, tension=0.7}{v1,v2}
			\fmf{plain}{v2,n1}
			\fmf{wiggly}{n1,n2,n3}
			\fmfv{d.s=circle, d.filled=shaded}{n2}
			\fmfv{label=$\overbrace{\phantom{phantom}}^{= 1}$, label.angle=90, label.dist=0.7pt}{n3}
			\fmf{plain}{n3,n4}
			\fmf{plain, tension=0.7}{n4,v3}
			\fmf{plain, tension=1}{o1,v3,o2}
			\fmfv{d.s=circle, d.filled=empty}{i1,i2,v2,n4,o1,o2}
		\end{fmfgraph*}		
	} \quad \\[20pt]
    {\bf b)} \quad $- 3 \cdot 3 \cdot$ \quad 
	\parbox{45mm}{
		\begin{fmfgraph*}(130,30)
			\fmfleft{i1}
			\fmfright{o1}
			\fmf{plain, tension=3.5}{i1,v1}	
			\fmf{plain, left=.7}{v1,v2,v1}
			\fmf{plain, tension=5}{v2,n1}
			\fmf{wiggly, tension=5}{n1,n2,n3}
			\fmfv{d.s=circle, d.filled=shaded}{n2}
			\fmfv{label=$\overbrace{\phantom{phantom}}^{= 1}$, label.angle=90, label.dist=0.7pt}{n3}
			\fmf{plain, tension=5}{n3,n4}
			\fmf{plain, tension=3.5}{n4,v3}
			\fmfv{d.s=circle, d.filled=empty}{n4}
			\fmf{plain, left=.7}{v3,o1,v3}
			\fmfv{d.s=circle, d.filled=empty}{i1,v2,o1}
		\end{fmfgraph*}		
	} \\[20pt]
	{\bf c)} \quad $- 3 \cdot 3 \cdot 1 \cdot$ \quad
	\parbox{50mm}{
		\begin{fmfgraph*}(145,30)
			\fmfleft{i1}
			\fmfright{o1}
			\fmf{plain, tension=3.5}{i1,v1}	
			\fmf{plain, left=.7}{v1,v2,v1}
			\fmf{plain, tension=5}{v2,n1}
			\fmf{wiggly, tension=5}{n1,n2,n3}
			\fmfv{d.s=circle, d.filled=shaded}{n2}
			\fmfv{label=$\underbrace{\phantom{He sieh mal das ist Phantomas!}}_\mathrm{additional\ non-cancelling\ diagram}$, label.angle=-90, label.dist=5pt}{n2}
			\fmf{plain, tension=5}{n3,v3}
			\fmfv{d.s=circle, d.filled=empty}{v3}
			\fmf{plain, left=.7}{v3,v4,v3}
			\fmf{plain,tension=3.5}{v4,o1}
			\fmfv{d.s=circle, d.filled=empty}{i1,v2,o1}
		\end{fmfgraph*}		
	} \\[5pt]
\end{fmffile}

\caption{Some diagrams generated by \prettyref{eq:add_reducible} at second
order $l=2$. Diagram a) and b) cancel reducible graphs b) and c)
in \prettyref{fig:Diagrams-connected-second-order}, respectively.\label{fig:reducible-diagrams-generated}}
\end{figure}

We observe that the diagram a) in \prettyref{fig:reducible-diagrams-generated}
cancels the diagram b) in \prettyref{fig:Diagrams-connected-second-order}.
This cancellation is of the ordinary type; the diagram b) in \prettyref{fig:Diagrams-connected-second-order}
is one-line irreducible in the original sense: its two components
are connected by a second order cumulant, which would be denoted by
a line $\Feyn{ff0}$ in the original graphical language of Feynman
diagrams; here denoted as $\Feyn{fcf}$.

An example of the more general cancellation appears between diagram
b) in \prettyref{fig:reducible-diagrams-generated} and c) in \prettyref{fig:Diagrams-connected-second-order}:
the two components in \prettyref{fig:Diagrams-connected-second-order}c)
are not connected by a second order cumulant, hence the diagram is
not one-line irreducible in the original sense, but it is reducible
in the sense we defined above. We see that it is canceled by \prettyref{fig:reducible-diagrams-generated}b),
consistent with the general proof. If we want to obtain the diagrams
of the next order, note that we have to keep reducible diagrams, because
the cancellation occurs only after setting $j=\Gamma_{0}^{\left(1\right)}\left(x^{\ast}\right)$.

The diagram \prettyref{fig:reducible-diagrams-generated}c) produced
by \prettyref{eq:add_reducible} cannot be generated by \prettyref{eq:add_connected},
because the connecting proper (effective) vertex $\Gamma_{0}^{(2)}=\Feyn{gpg}$
connects to two third order cumulants. It is generated in two ways;
with two $\Gamma^{\left(1\right)}\left(x^{\ast}\right)$-components
or only one of them. These two contributions have opposite signs,
but different prefactors, therefore they do not cancel (see also \prettyref{subsec:Taxonomy-of-reducible}
for details). We now derive the appearance of this latter diagram
by the application of the method using skeleton diagrams. The relevant
skeleton diagram is the one that appears at second order on the Taylor
expansion in the second line of \prettyref{eq:Taylor_Gamma_Skeleton}
\begin{eqnarray}
 &  & \frac{1}{2!}\,\Diagram{cfpfc}
\quad.\label{eq:skeleton_order_two}
\end{eqnarray}
We now instantiate this diagram with perturbative expansions of the
desired order, so that the resulting diagram is of order two: The
two point vertex function, by Dyson's equation \prettyref{eq:inversed_Dyson}
to lowest order is $\Gamma^{(2)}(x_{1})=\Gamma_{0}^{(2)}(x^{\ast})+\ldots$.
The leaves $-W_{V}^{(1)}$ attached to either side, to first order,
contain contributions of the form

\begin{fmffile}{First_order_cum_for_skeleton}	
	\begin{eqnarray}
	\parbox{30mm}{
		\begin{fmfgraph*}(70,30)
			\fmfpen{0.75thin}
			\fmfleft{i1}
			\fmfright{o1}
			\fmf{plain, tension=3.5}{i1,v3}
			\fmfv{d.s=circle, d.filled=empty}{v3}
			\fmf{plain, left=.7}{v3,v4,v3}
			\fmf{plain,tension=3.5}{v4,o1}
			\fmfv{d.s=circle, d.filled=empty}{o1}
		\end{fmfgraph*}		
	}+\dots,
	\end{eqnarray} 
\end{fmffile}where all elements on the right are cumulants of the unperturbed system
and bare interaction vertices. The combinatorial factor of this latter
sub-diagram is computed with the standard rules for connected diagrams,
which yields a factor $3$ from the possible ways of attaching the
first order cumulant to the interaction vertex. There is only a single
possibility to attach the additional external leg to the third order
cumulant and only a single way to join the elements in the skeleton
diagram in \prettyref{eq:skeleton_order_two}, besides the explicit
factor $1/2!$.

Finally, each remaining diagram must be interpreted algebraically
by steps (\prettyref{enu:assign-the-factor}) and (\prettyref{enu:j0_as_x_star}).
These steps are straight forward here. We note that this step is possible
at all in the technique using the skeleton diagrams, because all cumulants
appearing are $W_{0}^{(n)}(j_{0})=W_{0}^{(n)}(\Gamma_{0}^{(1)}(x^{\ast}))$
and in Dyson's equation \prettyref{eq:inversed_Dyson} in addition
only $\Gamma_{0}^{(2)}(x^{\ast})$ appears; so all elements, by assumption
of $S_{0}$ being the solvable theory, are known. There is yet another
observation that we can make in the framework of of this toy model,
namely that there are reducible non-standard diagrams that contribute
to $\Gamma_{V}$. One example is the following fourth-order cumulant
emerging from the $n=2$ term in the sum in \prettyref{eq:Vertex_Taylor_x1}
and using the first-order correction from \prettyref{eq:inversed_Dyson}
(with second order correction in the interaction for $W_{V}^{\left(2\right)}$):

\begin{fmffile}{Reducible_nonstandard}
\fmfset{thin}{0.75pt}
\fmfset{decor_size}{4mm}
	\parbox{50mm}{
		\begin{fmfgraph*}(335,30)
			\fmfleft{i1}
			\fmfright{o1}
			\fmf{plain, tension=3.5}{i1,v1}	
			\fmf{plain, left=.7}{v1,v2,v1}
			\fmf{plain, tension=5}{v2,n1}
			\fmf{wiggly, tension=5}{n1,n2,n3}
			\fmfv{d.s=circle, d.filled=shaded}{n2}
			\fmf{plain, tension=5}{n3,v3}
			\fmfv{d.s=circle, d.filled=empty}{v3}
			\fmf{plain, left=.7}{v3,v4,v3}
			\fmf{plain,tension=3.5}{v4,v5}
			\fmf{plain,tension=3.5}{v5,v6}	
			\fmf{plain, left=.7}{v6,v7,v6}
			\fmf{plain, tension=5}{v7,nn1}
			\fmf{wiggly, tension=5}{nn1,nn2,nn3}
			\fmfv{d.s=circle, d.filled=shaded}{nn2}
			\fmf{plain, tension=5}{nn3,v8}
			\fmfv{d.s=circle, d.filled=empty}{v3}
			\fmf{plain, left=.7}{v8,v9,v8}
			\fmf{plain,tension=3.5}{v9,o1}
			\fmfv{d.s=circle, d.filled=empty}{i1,v2,v5,v7,v8,o1}
			\fmfv{label=$\underbrace{\phantom{He sieh mal das ist Phantomas!}}_{\subseteq W_{V}^{\left(2\right)}}$, label.angle=-90, label.dist=5pt}{v5}
			\fmfv{label=$\underbrace{\phantom{He sieh mal} \mkern40mu}_{\subseteq W_{V}^{\left(1\right)}}$, label.angle=-68, label.dist=5pt}{v1}
			\fmfv{label=$\underbrace{\phantom{He sieh mal} \mkern40mu}_{\subseteq W_{V}^{\left(1\right)}}$, label.angle=-112, label.dist=5pt}{v9}
			\fmfv{label=$\underbrace{\phantom{He sieh}}_{=\Gamma_{0}^{\left(2\right)}}$, label.angle=-90, label.dist=5pt}{n2,nn2}
		\end{fmfgraph*}		
	} \\[5pt]
\end{fmffile}

\subsection{Application to the Ising model\label{sub:Application-to-Ising}}

We here illustrate the method on a simple system, the classical
Ising model with the action
\begin{eqnarray}
S(x) & = & \frac{\epsilon}{2}\,x^{\T}Jx+j^{\T}x,\label{eq:action_Ising}
\end{eqnarray}
where $x\in\{-1,1\}^{N}$ is a vector of Ising spins and $J$ a symmetric
matrix with $J_{ii}=0$. Note that there is no additional constraint
on $J$; it could, for example, be drawn from a random distribution,
as done in spin-glass models. In particular there is no restriction
to nearest-neighbour interactions. We are here interested in the weak
coupling limit with small $\epsilon$. 

There are essentially two different ways to arrive at an approximation
of the effective action $\Gamma(x^{\ast})$. The first approach represents
the pairwise coupling term as the result of a Gaussian average over
auxiliary variables (see e.g. \cite[Chapter 4.3 eq. 4.50a]{NegeleOrland98},
\cite[chapter (5.2), eq. (5.49)]{Vasiliev98}, \cite[eq. (4), (5)]{Sommers87_1268},
or \cite{Horwitz61}). This Hubbard-Stratonovich transform reduces
the calculation of $\Z$ to the summation of $n$-th moments of the
Gaussian employing Wick's theorem, weighted by the Taylor coefficients
of $V(x)=\ln\sum_{x_{i}}e^{j_{i}x_{i}}=\sum_{i}\,\ln\,2\cosh\,j_{i}$;
the latter play the role of vertices here. Consequently, standard
Feynman diagrammatic rules apply. This approach is only applicable
to this model because the interaction is pairwise; interactions of
higher order cannot be written by help of Gaussian auxiliary fields.

The second way, which we will follow here, considers as the solvable
part the single-spin term of the action
\begin{eqnarray}
S_{0}(x) & = & j^{\T}x;\qquad W_{0}(j)=\sum_{i}\,\ln\,2\,\cosh(j_{i}),\label{eq:S0_W0_Ising}
\end{eqnarray}
which is directly summable, yielding a cumulant-generating function
$W_{0}$ whose decomposition into a sum over individual sites shows
their statistical independence. For each $i$, the solvable theory
therefore has the infinitely many non-vanishing cumulants of a binary
variable. The interaction is in turn treated as the perturbation
\begin{eqnarray}
V(x) & = & \frac{\epsilon}{2}\,x^{\T}Jx.\label{eq:S_int_Ising}
\end{eqnarray}
This approach has been followed in quite a number of variations over
decades (see e.g. \cite[section 5.2, eq. 5.47]{Vasiliev98}, \cite{Bloch65,Vasiliev74,Bogolyubov1976,Plefka82_1971,Yedidia90,Georges91,Nakanishi97_8085,Tanaka98_2302},
\cite[chapter 3]{Opper01}); these methods agree inasmuch as they
all perform an expansion of $\Gamma$, where \prettyref{eq:S_int_Ising}
is treated as a perturbation and \prettyref{eq:S0_W0_Ising} as the
solvable part. In particular this approach is identical to Plefka's
method \cite{Plefka82_1971}.

Both routes of course lead to identical results. To second order in
$\epsilon$ one obtains the mean-field theory presented by Thouless,
Anderson, and Palmer (TAP) \cite{Thouless77_593} as a ``fait acompli'',
without proof, but mentioning a previous diagrammatic derivation.
Indeed, Vasiliev and Radzhabov (\cite{Vasiliev74,Vasiliev75}, summarized
in \cite[section 6.3.1, 6.3.2, 6.3.4]{Vasiliev98}) had derived this
result diagrammatically with the help of the analog to the Dyson Schwinger
equation for the effective action \cite{Dahmen67} before (see \cite[section 5.2]{Vasiliev98}
for a review of this method).

If one performed a perturbation expansion directly on the level of
$\Z$ \prettyref{eq:perturbation_general}
\begin{eqnarray}
\Z(j) & = & \exp(V(\partial_{j}))\,\exp(W_{0}(j))=\big\langle\sum_{n=0}^{\infty}\frac{1}{n!}(\frac{\epsilon}{2}\,x^{\T}Jx)^{n}\big\rangle_{0},\label{eq:brute_force}
\end{eqnarray}
with $\langle f(x)\rangle_{0}=\sum_{x}f(x)\,\exp(j^{\T}x),$ one would
quickly obtain unwieldy expressions; but most of these terms cancel
by the subsequent transformations $\Z\stackrel{\ln}{\rightarrow}W\stackrel{\mathcal{L}}{\rightarrow}\Gamma$
in the final result, making a direct expansion of $\Gamma$ desirable.
The calculation of $\Gamma$ at higher orders in $\epsilon$ by this
direct approach has prompted for computer algebra systems \cite{Nakanishi97_8085}.
Georges and Yedidia \cite{Yedidia90,Georges91} developed a sequence
of shortcuts and tricks for this problem by which they managed the
approximation up to forth order (reviewed in \cite[chapter 3]{Opper01}).
As mentioned, $\Gamma$ has been derived diagrammatically by Vasiliev
and Radzhabov \cite[eq. (21)]{Vasiliev74}. Using Dyson-Schwinger
equations, they explicitly presented all orders up to and including
the third.

We here chose this model to exemplify the application of the general
framework exposed in previous sections. For one, because it allows
the comparison to a wealth of methods developed over decades, as mentioned
above. And also because the Ising model has proven useful in many
other applications than micromagnetism, for example artificial neural
networks \cite{Hopfield82}. Especially the TAP-approximation and
its higher order corrections are employed to derive analytical approximations
\cite{Gabrie15} to contrastive divergence \cite{CarreiraPerpinan05},
a learning rule commonly employed to train restricted Boltzmann machines
\cite{Smolensky86,Hinton2006_504}. Another major field, in which
the Ising model is applied, are inference problems, where the inverse
Ising problem has to be solved, that is, the sources $h_{i}$ and
couplings $J_{ij}$ have to be computed for given means and pairwise
covariances. Depending on the problem, the spins represent activities
of neural units \cite{Tkacik06,Roudi09a,Hertz11}, active or inactive
genes in the case of gene regulatory networks or participants in financial
markets (see \cite{Nguyen17} for an excellent review of the inverse
Ising problem). The most natural quantity for this type of problem
would be the second Legendre transform, the usual entropy of the multi-variate
binary distribution for given first two moments,

 because the couplings and sources turn out to be the extrema of
this quantity. We present an iterative method to compute the second
Legendre transform in \prettyref{subsec:Second-Legendre-transform}.
Still, the first Legendre seems to be commonly used also for the inverse
problem \cite{Kappen98,Tanaka98_2302}. The reason for this might
be that calculating the second Legendre transform is technically
challenging; nevertheless, specific approximations exist that are
valid for small correlations, first computed by a technique specialized
to the Ising model \cite{Sessak09}, later obtained by techniques
borrowed from the functional renormalization group \cite{Jacquin16},
using the Wetterich equation \cite{WETTERICH93_90}.. An extension
to more than pairwise coupling \prettyref{eq:S_int_Ising} is desirable
in these fields \cite{Roudi09,Schneidman06_1007,Tkacik14_e1003408}.
In particular the Hubbard-Stratonovich method mentioned above would
not work in these cases, while the here proposed approach does.

Performing the here presented diagrammatic expansion, we observe a
drastic decrease of computations and their transparent organization
by diagrams. The calculation up to fourth order indeed takes only
minor effort.

We start with the unperturbed theory, given by \prettyref{eq:S0_W0_Ising},
in which all spins decouple and the cumulants up to fourth order read
\begin{eqnarray*}
\llangle x_{i}\rrangle_{0} & = & \left\langle x_{i}\right\rangle _{0}=m_{i}:=\tanh\left(j_{i}\right)\\
\llangle x_{i}x_{j}\rrangle_{0} & = & \delta_{ij}\left(1-m_{i}^{2}\right)\\
\llangle x_{i}x_{j}x_{k}\rrangle_{0} & = & -\delta_{ij}\delta_{jk}2m_{i}\left(1-m_{i}^{2}\right)\\
\llangle x_{i}x_{j}x_{k}x_{l}\rrangle_{0} & = & -\delta_{ij}\delta_{jk}\delta_{kl}2\left(1-3m_{i}^{2}\right)\left(1-m_{i}^{2}\right).
\end{eqnarray*}
To zeroth order the Legendre transform of $W_{0}$ is the entropy
of a binary variable
\begin{eqnarray}
\Gamma_{0}(m) & = & -\sum_{i}\frac{1+m_{i}}{2}\,\ln\left(\frac{1+m_{i}}{2}\right)+\frac{1-m_{i}}{2}\,\ln\left(\frac{1-m_{i}}{2}\right),\label{eq:Gamma_0}
\end{eqnarray}
where $(1\pm m_{i})/2$ are the probabilities for $x_{i}=\pm1$. The
result \prettyref{eq:Gamma_0} follows from step (\prettyref{enu:Calculate-Gamma0})
of the algorithm by a short calculation from \prettyref{eq:S0_W0_Ising};
the form is moreover clear, because the distribution \prettyref{eq:S0_W0_Ising}
maximizes the entropy and the Legendre transform to $\Gamma$ just
fixes $j$ so that the mean is $\langle x_{i}\rangle=m_{i}$.

To obtain corrections to \prettyref{eq:Gamma_0} we represent the
$n$-th cumulant by an empty circle with $n$ legs and the $n$-th
derivative of $\Gamma_{0}$ by a hatched circle with $n$ legs. An
interaction $J_{ij}$ is denoted by an edge:

\begin{fmffile}{Example_feyn}	
	\begin{eqnarray}
		\parbox{25mm}{
			\begin{fmfgraph*}(25,25)
				\fmfpen{0.5thin}
				\fmftop{o1,o2,o3}
				\fmfbottom{u1,u2,u3}
				\fmf{plain}{u1,o2}
				\fmf{plain}{u3,o2}
				\fmfv{label=$i$, label.angle=90, label.dist=-10pt}{u1}
				\fmfv{label=$j$, label.angle=90, label.dist=-10pt}{u3}
			\end{fmfgraph*}
		}  \mkern-72mu := J_{ij},
		\parbox{25mm}{
			\begin{fmfgraph*}(25,25)
				\fmfpen{0.5thin}
				\fmftop{o1,o2,o3,o4,o5}
				\fmfbottom{u1,u2,u3,u4,u5}
				\fmf{phantom}{u1,m1,o1}
				\fmf{phantom}{u3,m3,o3}
				\fmf{phantom}{u5,m5,o5}
				\fmf{plain}{m3,o4}
				\fmf{plain}{m3,o5}
				\fmf{plain,tension=0}{m3,m5}
				\fmf{phantom, tension=10}{m1,m3}
				\fmf{plain,tension=2.5}{m3,u5}
				\fmfv{decor.shape=circle,decor.filled=empty, decor.size=6.5thin}{m3}
				\fmfv{label=$i_{1}$, label.angle=90, label.dist=4.5pt}{o4}
				\fmfv{label=$i_{2}$, label.angle=60, label.dist=2.pt}{o5}
				\fmfv{label=$i_{3}$, label.angle=20, label.dist=1.5pt}{m5}
				\fmfv{label=$...$, label.angle=0, label.dist=1.5pt}{u5}
			\end{fmfgraph*}
		}  \mkern-63mu := \llangle x_{i_{1}}x_{i_{2}}x_{i_{3}}...\rrangle,
			\; \parbox{25mm}{
			\begin{fmfgraph*}(25,25)
				\fmfpen{0.5thin}
				\fmftop{o1,o2,o3,o4,o5}
				\fmfbottom{u1,u2,u3,u4,u5}
				\fmf{phantom}{u1,m1,o1}
				\fmf{phantom}{u3,m3,o3}
				\fmf{phantom}{u5,m5,o5}
				\fmf{wiggly}{m3,o4}
				\fmf{wiggly}{m3,o5}
				\fmf{wiggly, tension=0}{m3,m5}
				\fmf{phantom, tension=10}{m1,m3}
				\fmf{wiggly,tension=2.5}{m3,u5}
				\fmfv{decor.shape=circle,decor.filled=shaded, decor.size=6.5thin}{m3}
				\fmfv{label=$i_{1}$, label.angle=90, label.dist=4.5pt}{o4}
				\fmfv{label=$i_{2}$, label.angle=60, label.dist=2.pt}{o5}
				\fmfv{label=$i_{3}$, label.angle=20, label.dist=1.5pt}{m5}
				\fmfv{label=$...$, label.angle=0, label.dist=1.5pt}{u5}
			\end{fmfgraph*}
		}  \mkern-63mu := \Gamma^{\left(n\right)}_{i_{1},i_{2},i_{3},...}.\\ \nonumber
	\end{eqnarray}
\end{fmffile}To evaluate the diagrams, we will only need the second derivative
$\Gamma_{0}^{\left(2\right)}$, either directly given by differentiating
\prettyref{eq:Gamma_0} twice or by using the relation $W_{0}^{\left(2\right)}\Gamma_{0}^{\left(2\right)}=1$.
Both yields $\Gamma_{0,ij}^{\left(2\right)}=\frac{\delta_{ij}}{1-m_{i}^{2}}$.
Within this language, the perturbative corrections up to the third
order to be added to \prettyref{eq:Gamma_0} are readily constructed
from steps \ref{enu:recursion}-\ref{enu:j0_as_x_star} at the end
of \prettyref{sub:Graphical-rules-for-Gamma}:

\begin{fmffile}{Gamma_first_three}	
	\begin{eqnarray}
		\parbox{25mm}{
			\begin{fmfgraph*}(25,25)
				\fmfpen{0.5thin}
				\fmftop{o1,o2,o3}
				\fmfbottom{u1,u2,u3}
				\fmf{plain}{u1,o2}
				\fmf{plain}{u3,o2}
				\fmfv{decor.shape=circle,decor.filled=empty, decor.size=6.5thin}{u1,u3}
			\end{fmfgraph*}
		}   &\mkern-60mu= \frac{1}{2}\sum_{i\neq j} J_{ij}m_{i}m_{j}&\\
       \mkern-40mu \parbox{25mm}{
			\begin{fmfgraph*}(75,25)
				\fmfpen{0.5thin}
				\fmftop{o1,o2,o3,o4,o5}
				\fmfbottom{u1,u2,u3,u4,u5}
				\fmf{phantom}{u1,v1,o3}
				\fmf{plain}{v1,o3}
				\fmf{phantom}{o1,v1,u3}
				\fmf{plain}{v1,u3}
				\fmf{phantom}{u3,v2,o5}
				\fmf{plain}{u3,v2}
				\fmf{phantom}{o3,v2,u5}
				\fmf{plain}{o3,v2}
				\fmfv{decor.shape=circle,decor.filled=empty, decor.size=6.5thin}{v1,v2}
			\end{fmfgraph*}
			}
		& \mkern-60mu = \frac{1}{2!2^{2}} 2 \sum_{i\neq j} J_{ij}^{2}\left(1-m_{i}^{2}\right)\left(1-m_{j}^{2}\right)  \label{Def_second_oder_diagram}&\\
		\mkern-40mu \parbox{25mm}{
			\begin{fmfgraph*}(75,25)
				\fmfpen{0.5thin}
				\fmftop{o1,o2,o3}
				\fmfbottom{u1,u2,u3}
				\fmf{phantom,tension=100}{u1,dl,v1,o2}
				\fmf{plain}{dul,v1,o2}
				\fmf{phantom,tension=100}{u3,dr,v2,o2}
				\fmf{plain}{dur,v2,o2}
				\fmf{phantom}{u1,dul,u2,dur,u3}
				\fmf{plain}{dul,u2,dur}
				\fmf{phantom,tension=0.5}{dl,dum,dr}
				\fmf{phantom,tension=1}{dum,u2}
				\fmfv{decor.shape=circle,decor.filled=empty, decor.size=6.5thin}{v1,v2,u2}
			\end{fmfgraph*}
		}		
		& \mkern-60mu = \frac{1}{3!2^{3}} 2^{3} \sum_{i\neq j\neq k\neq i} J_{ij} J_{jk}J_{ki}\left(1-m_{i}^{2}\right)\left(1-m_{j}^{2}\right)\left(1-m_{k}^{2}\right)\label{Def_third_oder_ring_diagram}&\\
		\parbox{25mm}{
			\begin{fmfgraph*}(25,25)
				\fmfpen{0.5thin}
				\fmftop{o1,o2,o3}
				\fmfbottom{u1,u2,u3}
				\fmf{phantom}{u2,du,do,o2}
				\fmf{phantom,tension=2}{du,do,o2}
				\fmf{plain}{u1,o2}
				\fmf{plain}{u3,o2}
				\fmf{plain}{u1,do}
				\fmf{plain}{u3,do}
				\fmf{plain}{u1,du}
				\fmf{plain}{u3,du}
				\fmfv{decor.shape=circle,decor.filled=empty, decor.size=6.5thin}{u1,u3}
			\end{fmfgraph*}
		}
		& \mkern-60mu = \frac{1}{3!2^{3}} 2^{2} \sum_{i\neq j} J_{ij}^{3} \left(-2m_{i}\right)\left(1-m_{i}^{2}\right)\left(-2m_{j}\right)\left(1-m_{j}^{2}\right).\label{Def_third_oder_Citroen_diagram}&
	\end{eqnarray}
\end{fmffile}

For the third order diagrams, the symmetry factors $2^{3}$ and $2^{2}$
are noted separately in front of the respective terms. For all diagrams,
they are determined as usual for Feynman diagrams (see \prettyref{app:App_Calc_of_sym_factors}).
Until this order only the first sort of ordinary diagrams composed
of vertices and cumulants contribute. All diagrams are irreducible
in the general sense defined above. 

Going to fourth order, we first consider the four diagrams that are
formed out of vertices and cumulants alone. These would also contribute
to an expansion of $W$.\begin{fmffile}{Gamma_fourth_ordinary}	
	\begin{eqnarray}
		\mkern-150mu \parbox{25mm}{
			\begin{fmfgraph*}(75,25)
				\fmfpen{0.5thin}
				\fmftop{o1,o2,o3}
				\fmfbottom{u1,u2,u3}
				\fmf{phantom}{u1,dul,dml,v1,o2}
				\fmf{plain}{dml,v1,o2}
				\fmf{phantom}{u2,v2,dml,dol,o1}
				\fmf{plain}{u2,v2,dml}
				\fmf{phantom}{u2,v3,dmr,dor,o3}
				\fmf{plain}{u2,v3,dmr}
				\fmf{phantom}{u3,dur,dmr,v4,o2}
				\fmf{plain}{dmr,v4,o2}
				\fmfv{decor.shape=circle,decor.filled=empty, decor.size=6.5thin}{v1,v2,v3,v4}
			\end{fmfgraph*}
		}			
		& \mkern-30mu = \frac{S_{\mathrm{R}}}{4!2^{4}}\sum_{i\neq j\neq k \neq l\neq i} J_{ij}J_{jk}J_{kl}J_{li} \left(1-m_{i}^{2}\right)\left(1-m_{j}^{2}\right)\left(1-m_{k}^{2}\right)\left(1-m_{l}^{2}\right)  \label{Fourth_order_ring}&\\
			\mkern-150mu\parbox{25mm}{
			\begin{fmfgraph*}(75,25)
				\fmfpen{0.5thin}
				\fmftop{o1,o2,o3}
				\fmfbottom{u1,u2,u3}
				\fmf{phantom,tension=100}{u1,dl,v1,o2}
				\fmf{plain}{dul,v1,o2}
				\fmf{phantom,tension=100}{u3,dr,v2,o2}
				\fmf{plain}{dur,v2,o2}
				\fmf{phantom}{u1,dul,u2,dur,u3}
				\fmf{plain}{dul,u2,dur}
				\fmf{phantom,tension=0.5}{dl,dum,dr}
				\fmf{phantom,tension=1}{dum,u2}
				\fmf{plain}{v2,dum}
				\fmf{plain}{v1,dum}
				\fmfv{decor.shape=circle,decor.filled=empty, decor.size=6.5thin}{v1,v2,u2}
			\end{fmfgraph*}
		}
		&\mkern-30mu =\frac{S_{\mathrm{TM}}}{4!2^{4}}\sum_{i\neq j\neq k\neq i} J_{ik}J_{ij}^2J_{kj} \left(-2m_{i}\right)\left(1-m_{i}^{2}\right)\left(-2m_{j}\right)\left(1-m_{j}^{2}\right)\left(1-m_{k}^{2}\right)\label{Fourth_order_TM}&\\
		\mkern-150mu\parbox{25mm}{
			\begin{fmfgraph*}(75,25)
				\fmfpen{0.5thin}
				\fmftop{o0,o1,o2,o3,o4,o5,o6}
				\fmfbottom{u0,u1,u2,u3,u4,u5,u6}
				\fmf{phantom}{u1,v1,o3}
				\fmf{plain}{v1,o3}
				\fmf{phantom}{o1,v1,u3}
				\fmf{plain}{v1,u3}
				\fmf{phantom}{u3,v2,o5}
				\fmf{plain}{u3,v2}
				\fmf{phantom}{o3,v2,u5}
				\fmf{plain}{o3,v2}
				\fmf{phantom}{u0,v1,dummy1,o4}
				\fmf{plain}{v1,dummy1}
				\fmf{phantom}{u4,dummy2,v1,o0}
				\fmf{plain}{dummy2,v1}
				\fmf{phantom}{u2,dummy2,v2,o6}
				\fmf{plain}{dummy2,v2}
				\fmf{phantom}{u6,v2,dummy1,o2}
				\fmf{plain}{v2,dummy1}
				\fmfv{decor.shape=circle,decor.filled=empty, decor.size=6.5thin}{v1,v2}
			\end{fmfgraph*}
		}
		& \mkern-30mu = \frac{S_{\mathrm{dC}}}{4!2^{4}}\sum_{i\neq j} J_{ij}^{4}\left(-2\right)\left(1-3m_{i}^{2}\right)\left(1-m_{i}^{2}\right)\left(-2\right)\left(1-3m_{j}^{2}\right)\left(1-m_{j}^{2}\right).\label{Fourth_order_Citroen}&
	\end{eqnarray}
\end{fmffile}

\selectlanguage{english}%

\selectlanguage{american}%
In the appendix, \prettyref{app:App_Calc_of_sym_factors}, we show
that the symmetry factors are given by $\SR=48$, $\ST=96$ and $\SdC=8$.
The fourth standard diagram is shown below. It is similar to the first
 and second non-standard contribution. To indicate the origin of the
different contributions, we denote sub-graphs originating from $\Gamma^{\left(1\right)}$
by filled circles - they are, however, translated in the same way
as the empty ones:

\selectlanguage{english}%
\begin{fmffile}{Gans}
	\begin{eqnarray}
    \label{Fourth_order_extraordinary}
			& \underbrace{\mkern-18mu
				\parbox{20mm}{
				\begin{fmfgraph*}(75,25)
					\fmfpen{0.5thin}
					\fmftop{o1,o2,o3,o4,o5,o6,o7}
					\fmfbottom{u1,u2,u3,u4,u5,u6,u7}
					\fmfright{r1}
					\fmfleft{l1}
					\fmf{phantom}{v3,r1}
					\fmf{phantom}{l1,v1}
					\fmf{phantom}{u1,v1,o3}
					\fmf{plain}{v1,o3}
					\fmf{phantom}{o1,v1,u3}
					\fmf{plain}{v1,u3}
					\fmf{plain}{u3,v2,o5}
					\fmf{plain}{o3,v2,u5}
					\fmf{phantom}{u5,v3,o7}
					\fmf{plain}{u5,v3}
					\fmf{phantom}{o5,v3,u7}
					\fmf{plain}{o5,v3}
					\fmfv{decor.shape=circle,decor.filled=empty, decor.size=6.5thin}{v1,v2,v3}
				\end{fmfgraph*}
				}\quad}_{a}
				-
				\underbrace{\mkern-18mu
				\parbox{20mm}{
				\begin{fmfgraph*}(100,25)
					\fmfpen{0.5thin}
					\fmftop{o1,o2,o3,o4,o5,o6,o7,o8,o9}
					\fmfbottom{u1,u2,u3,u4,u5,u6,u7,u8,u9}
					\fmfright{r1}
					\fmfleft{l1}
					\fmf{phantom}{v3,r1}
					\fmf{phantom}{l1,v1}
					\fmf{phantom}{u1,v1,o3}
					\fmf{plain}{v1,o3}
					\fmf{phantom}{o1,v1,u3}
					\fmf{plain}{v1,u3}
					\fmf{phantom}{u3,v2,o5}
					\fmf{plain}{u3,v2}
					\fmf{phantom}{o3,v2,u5}
					\fmf{plain}{o3,v2}
					\fmf{phantom}{u5,v4,o7}
					\fmf{plain}{v4,o7}			
					\fmf{phantom}{u7,v4,o5}
					\fmf{plain}{u7,v4}
					\fmf{phantom}{u7,v3,o9}
					\fmf{plain}{u7,v3}
					\fmf{phantom}{o7,v3,u9}
					\fmf{plain}{o7,v3}
					\fmf{plain}{v2,g1}
					\fmf{wiggly}{g1,vv1}
					\fmf{wiggly}{vv1,g2}
					\fmf{plain}{g2,v4}
					\fmfv{decor.shape=circle,decor.filled=empty, decor.size=6.5thin}{v1,v2}
					\fmfv{decor.shape=circle,decor.filled=full, decor.size=6.5thin}{v3,v4}
					\fmfv{decor.shape=circle,decor.filled=shaded, decor.size=6.5thin}{vv1}
				\end{fmfgraph*}
				}
				\quad
				\quad
				\quad}_{b}
                \;
				+
				\underbrace{\mkern-18mu
				\parbox{20mm}{
				\begin{fmfgraph*}(125,25)
					\fmfpen{0.5thin}
					\fmftop{o1,o2,o3,o4,o5,o6,o7,o8,o9,o10,o11}
					\fmfbottom{u1,u2,u3,u4,u5,u6,u7,u8,u9,u10,u11}
					\fmfright{r1}
					\fmfleft{l1}
					\fmf{phantom}{v3,r1}
					\fmf{phantom}{l1,v1}
					\fmf{phantom}{u1,v1,o3}
					\fmf{plain}{v1,o3}
					\fmf{phantom}{o1,v1,u3}
					\fmf{plain}{v1,u3}
					\fmf{phantom}{u3,v2,o5}
					\fmf{plain}{u3,v2}
					\fmf{phantom}{o3,v2,u5}
					\fmf{plain}{o3,v2}
					\fmf{phantom}{u7,v4,o9}
					\fmf{plain}{v4,o9}			
					\fmf{phantom}{u9,v4,o7}
					\fmf{plain}{u9,v4}
					\fmf{phantom}{u9,v3,o11}
					\fmf{plain}{u9,v3}
					\fmf{phantom}{o9,v3,u11}
					\fmf{plain}{o9,v3}
					\fmf{plain}{v2,g1}
					\fmf{wiggly}{g1,vv1}
					\fmf{wiggly}{vv1,g2}
					\fmf{plain}{g2,v5}
					\fmf{plain}{v5,g3}
					\fmf{wiggly}{g3,vv2}
					\fmf{wiggly}{vv2,g4}
					\fmf{plain}{g4,v4}
					\fmfv{decor.shape=circle,decor.filled=empty, decor.size=6.5thin}{v5}
					\fmfv{decor.shape=circle,decor.filled=full, decor.size=6.5thin}{v1,v2,v3,v4}
					\fmfv{decor.shape=circle,decor.filled=shaded, decor.size=6.5thin}{vv1,vv2}
					\fmfv{label=$\overbrace{ \enskip\quad\quad }^{=1}$, label.angle=90, label.dist=4pt}{g3}
				\end{fmfgraph*}
				}
				\quad\quad\quad\quad\quad\quad}_{c}&\\ \nonumber
			&=\frac{1}{2^{4}}\sum_{i\neq j\neq k} J_{ij}^{2}J_{jk}^{2}\\ \nonumber
& \times \left[\frac{S_{a}}{4!}\left(1-m_{i}^{2}\right)\left(-2\right)\left(1-m_{j}^{2}\right)\left(1-3m_{j}^{2}\right)\left(1-m_{k}^{2}\right)\right.& \\  \nonumber
			 & - \left(1-m_{i}^{2}\right) \frac{\left(-2m_{j}\right)\left(1-m_{j}^{2}\right)\left(-2m_{j}\right)\left(1-m_{j}^{2}\right)}{\left(1-m_{j}^{2}\right)} \left(1-m_{k}^{2}\right) \left. \left(\frac{S_{b}}{3!}-\frac{S_{c}}{2!}\right) \right] & \\  \nonumber
	\end{eqnarray}	
\end{fmffile}

\selectlanguage{american}%
With the symmetry factors given by $S_{a}=48$, $S_{b}=24$ and $S_{c}=4$
(see appendix, sec. \prettyref{app:App_Calc_of_sym_factors} for the
derivation), we can add up the contributions of the three diagrams,
which gives, leaving out factors that are equal in all diagrams for
simplicity:
\begin{eqnarray*}
\frac{1}{4!}48\left(-2\right)\left(1-3m_{j}^{2}\right)+\left(-\frac{1}{1!2!}8+\frac{1}{2}4\right)4m_{j}^{2} & = & -4\left(1-m_{j}^{2}\right).
\end{eqnarray*}
The last term looks like a second cumulant and, interestingly, this
contribution is indeed exactly canceled by the contribution of the
ring-diagram of fourth order (\ref{Fourth_order_ring}) in the case
that exactly one pair of indices, which belong to cumulants represented
at opposite sites of the ring, are equal. 

This has to be so, as shown by Vasiliev and Radzhabov \cite{Vasiliev75},
because, according to their terminology, diagram $a$ is a ``nonstar
graph'', that is, not a star graph - it decomposes into two parts
by removing a cumulant (a ``vertex'' in their words). Note that
the star graph property implies irreducibility, but not vice versa.
As Vasiliev and Radzhabov have shown, only star diagrams contribute
to the effective action of the Ising model, if we introduce so called
compensating graphs. These graphs are constructed as follows: Take
all star graphs of a certain order and draw contractions (in \cite{Vasiliev75}
depicted by dashed lines) in all possible ways between circles that
are not connected by an interaction and therefore represent indices
that could take the same value (keep in mind that $J_{ii}=0$). If
two or more circles are contracted, they are considered as a single
circle, are associated with a common index and are translated as the
product of all cumulants associated to this point. Then substract
all compensating graphs that are non-star-graphs and sum up original
and compensating graphs \cite[eq. (8)]{Vasiliev75}.  We may check
that our result is consistent with this rule. In the ring-diagram
(\ref{Fourth_order_ring}), the only contraction that leads to a non-star
graph is the one that identifies the indices of two opposite circles.
This contribution therefore cancels from the final expression for
the effective action, analogous to the summed up contribution of the
diagrams $a$, $b$, and $c$. In our framework not only the double
Citro{\"e}n diagram (\ref{Fourth_order_Citroen}) contributes to
the sum over only two unequal indices, but also the ring diagram in
the case that the indices of pairwise opposite circles and the diagrams
$a$, \textbf{$b$} and $c$ in the case that the indices of the outmost
circles are equal (see \prettyref{app:App_Calc_of_sym_factors} for
details). In summary, the general method that we present here is consistent
with the specific result known for the Ising model.

In total, we obtain the following expression for the effective action
in an Ising model with coupling $J_{ij}$:
\begin{eqnarray}
 & -\Gamma\left(\boldsymbol{m}\right)\label{eq:Effective_action_Ising}\\
= & -\sum_{i}\frac{1+m_{i}}{2}\ln\left(\frac{1+m_{i}}{2}\right)+\frac{1-m_{i}}{2}\ln\left(\frac{1-m_{i}}{2}\right)\nonumber \\
+ & \frac{1}{2}\epsilon\sum_{i\neq j}J_{ij}m_{i}m_{j}+\frac{1}{4}\epsilon^{2}\sum_{i\neq j}J_{ij}^{2}\left(1-m_{i}^{2}\right)\left(1-m_{j}^{2}\right)\nonumber \\
+ & \frac{1}{6}\epsilon^{3}\sum_{i\neq j\neq k\neq i}J_{ij}J_{jk}J_{ki}\left(1-m_{i}^{2}\right)\left(1-m_{j}^{2}\right)\left(1-m_{k}^{2}\right)\nonumber \\
+ & \frac{1}{3}\epsilon^{3}\sum_{i\neq j}J_{ij}^{3}m_{i}\left(1-m_{i}^{2}\right)m_{j}\left(1-m_{j}^{2}\right)\nonumber \\
+ & \frac{1}{8}\epsilon^{4}\sum_{\overset{i\neq j\neq k\neq l\neq i,}{i\neq k,j\neq l}}J_{ij}J_{jk}J_{kl}J_{li}\left(1-m_{i}^{2}\right)\left(1-m_{j}^{2}\right)\left(1-m_{k}^{2}\right)\left(1-m_{l}^{2}\right)\label{eq:fourth_order_I}\\
+ & \epsilon^{4}\sum_{i\neq j\neq k\neq i}J_{ik}J_{ij}^{2}J_{kj}m_{i}\left(1-m_{i}^{2}\right)m_{j}\left(1-m_{j}^{2}\right)\left(1-m_{k}^{2}\right)\label{eq:fourth_order_II}\\
- & \frac{1}{24}\epsilon^{4}\sum_{i\neq j}J_{ij}^{4}\left(1-m_{i}^{2}\right)\left(1-m_{j}^{2}\right)\left(1+3m_{i}^{2}+3m_{j}^{2}-15m_{i}^{2}m_{j}^{2}\right)\nonumber 
\end{eqnarray}

This calculation reproduces (with $\epsilon=\beta$) eq. (11) of Georges
and Yedidia \cite{Georges91} and is in line with equations (9), (10),
(14), and (15) in Nakanishi and Takayama \cite{Nakanishi97_8085}
\footnote{In comparing the expressions to these works, note the different sum
conventions: Georges et al. and Nakanishi et al. are only summing
over those tuples of distinct indices that lead to different terms,
while we allow multiple occurrence of terms already in the definition
of the action \prettyref{eq:action_Ising} and correct this by using
$\frac{J_{ij}}{2}$ instead of $J_{ij}$ as the interaction. Note
that the former sum convention in the second-to-last line of eq. (11)
in Georges et al. amounts to a summation over all tuples of three
different indices and all even permutations thereof, whereas in eq.
(15) of Nakanishi et al, these permutations are written out explicitly,
therefore the summation is only over the tuples with three different
indices. Therefore, even if these two expressions seem to disagree
by a factor $3$ at first sight, they do not, because, depending on
the form of the sum, the interpretation of their sum convention changes.
We thank Adam Ran\c con for clarifying this summation convention.} and the equations (B2)-(B5) in Jacquin and Ran\c con \cite{Jacquin16},
who derived this result using the Wetterich equation \cite{WETTERICH93_90}.

\section{Discussion}

We presented a systematic diagrammatic scheme to calculate the effective
action for any problem that can be decomposed into a solvable part
with known correlators and a perturbing part. We have proved that
corrections are composed of two types of diagrams:
\begin{itemize}
\item irreducible diagrams in a more general sense than in the Gaussian
case: those that cannot be decomposed by detaching a single leg of
a vertex;
\item diagrams of special form that are neither contained in the perturbation
expansions of $\Z$ nor $W$; they are composed of sub-graphs that
are connected by either a vertex $\Gamma_{0}^{(n)},\ n\geq2$ that,
by at least two legs, connects to a third or higher order bare cumulant.
This set of diagrams is found by instantiating a known sequence of
skeleton diagrams to the desired order in perturbation theory.
\end{itemize}
The appearance of the latter diagrams can be regarded a as generalization
of the amputation in the Gaussian case, because the inverse propagator
$\Gamma_{0}^{(2)}$ may attach to cumulants of the solvable problem
at any order, not only to second order cumulants; only in the latter
case it ``amputates'' the lines.

The presented inductive proof in addition yields an iterative equation
which allows the algebraic construction of all graphs and their combinatorial
factors from elementary rules of calculus.

One may wonder why we derived the recursion for $\Gamma$ based on
the ideas of the proof of the linked cluster theorem (see \prettyref{app:Perturbative-cumulant-expansion},
extending the proof in \cite[Sec. 6.1.1]{ZinnJustin96} to the non-Gaussian
case). It might seem more direct to modify the corresponding proof
of one-line irreducibility to the non-Gaussian setting considered
here. The latter, however, appeared impossible to us: The elegant
proof by Zinn-Justin \cite[section 6.5]{ZinnJustin96} rests on the
assumption that the underlying theory is Gaussian; in their eq. 6.59
each line is disconnected in all possible ways and the result is shown
to remain connected. The proof hence requires that the only connecting
elements of the bare theory be lines; this is precisely the restriction
we lift here.

The proof by Weinberg \cite[section 16.1]{Weinberg05_II} shows elegantly
that $W$ decomposes into tree graphs, whose vertices - which one
could call ``effective vertices'' in this case - are generated by
$\Gamma$. This statement remains of course true also in the non-Gaussian
case, because $W\stackrel{\mathcal{L}}{\leftrightarrow}\Gamma$ form
a pair of Legendre transforms. We use this fact to derive the decomposition
into skeleton diagrams. This decomposition also follows directly from
the reciprocity relation $1=\Gamma^{(2)}W^{(2)}$ of the Hessians
\cite[section 6.2]{ZinnJustin96}. The connecting elements in these
trees are the full propagators $W^{(2)}$, which Dyson's equation
expresses in terms of bare propagators and the self energy. The
reciprocity relation implies the recursion for the self-energy $\Gamma_{V}^{(2)}=-(\Gamma_{0}^{(2)}+\Gamma_{V}^{(2)})\,W_{V}^{(2)}\,\Gamma_{0}^{(2)}.$
Since $W_{V}$ is composed of all connected graphs, including those
containing higher order cumulants of $W_{0}$, we see that the terms
$\cdots\Gamma_{0}^{(2)}W_{V}\Gamma_{0}^{(2)}\cdots$ appearing in
the iteration produce those unusual terms of the form $W_{0}^{(n\ge2)}\Gamma_{0}^{(2)}W_{0}^{(m\ge2)}$
that we found in the general expansion (Eq. \ref{eq:reducible_general-1}).

An alternative approach is the analog of the Dyson Schwinger equation
for the effective action \cite{Dahmen67,Vasiliev73,Vasiliev74a},
reviewed in \cite[sec. 1.8, 6.2]{Vasiliev98}. It leads to equations
of motion for $\Gamma$ that enable an iterative expansion, for example
in the interaction strength. An iterative solution leads to a proof
of the 1PI property in the Gaussian case \cite[sec. 1.8.3]{Vasiliev98}.
We believe that the latter approach could be generalized to obtain
the same results as presented here. The equation of motion can be
derived from the invariance of the integral measure with respect to
translations \cite[sec. 1.8.3]{Vasiliev98}, but also extends to problems
on discrete state spaces, such as the Ising model \cite{Vasiliev74,Vasiliev75,Bogolyubov1976,Vasiliev98}.
From this equation of motion, eq. (6.136) in \cite{Vasiliev98} is
derived, similar to our \prettyref{eq:Gamma_skeleton_final} with
the difference that Vasiliev expands in terms of unperturbed cumulants
and not full vertices. Probably, both equations are suitable starting
points to rederive our results in \prettyref{sub:Graphical-rules-for-Gamma}.
The Feynman rules would then be deduced from this step. Possibly these
approaches could render the taxonomy of non-standard diagrams clearer.
However, it will unlikely be as simple as in the Gaussian case because,
as pointed out, reducible non-standard diagrams do not necessarily
cancel. We leave this question for future work. In any case, more
refined diagrammatic rules previously had to be obtained in a case
by case manner so far, depending on the problem at hand (compare also
the last paragraph of \cite[chap. 6.3.1.]{Vasiliev98}). The algorithm
presented in \prettyref{sub:Graphical-rules-for-Gamma}, in contrast,
is the same for any model.

One notes that eq. (21) in the work by Vasiliev and Radzhabov \cite{Vasiliev74}
is identical to the third order derivation recovered much later by
other means \cite{Plefka82_1971,Georges91,Nakanishi97_8085,Tanaka98_2302};
in particular the result already includes the TAP approximation \cite{Thouless77_593},
if only the first three terms are considered. In principle, these
early works \cite{Vasiliev74,Vasiliev75} also derive the Feynman
rules for the Ising model; however, without giving any concrete expression
for orders higher than three. In contrast to these results, the set
of Feynman rules that we present here are applicable to general non-Gaussian
theories. In case of the Ising model these had been sought for some
time (see \cite[p. 28]{Opper01}).

We hope that the presented technique may prove useful in finding new
approximations around known limiting cases. Examples may include expansions
of the Hubbard model around the atomic limit \cite{Georges91}. An
extension of our theory to higher order Legendre transforms, as broadly
discussed in \cite{Vasiliev98}, could lead to a diagrammatic formulation
of the results derived in the context of the inverse Ising problem
\cite{Sessak09,Jacquin16} and to an alternative field-theoretic formulation
of the extended Plefka-expansion for stochastic systems, that has
recently been developed \cite{Bravi16}. The appendix \prettyref{subsec:Second-Legendre-transform}
presents an iterative algorithm as a first step towards this goal,
an iterative equation to compute corrections to the second Legendre
transform. Further work is required to derive a set of Feynman rules
from this equation.

The application of the here presented method is possible whenever
a model admits a closed-form solution. An interesting regime of application
may therefore be spherical models \cite{Berlin52_821}. In the thermodynamic
limit, the free energy, the cumulant-generating function of the model,
is known. Extensions of the spherical model that include four point
coupling terms for example appear in the field of random lasers \cite{Antenucci15_053816}.
If the quartic term is small compared to the quadratic term, the here
proposed method could be applied to obtain approximate self-consistency
equations. Such quartic spherical models, moreover, appear in inverse
problems of diverse systems \cite{Marruzzo17}. Future work is needed
to see if the perturbative results offered by the current work may
help at obtaining approximate solutions to such inverse problems. 

We also note that the procedure as presented in \prettyref{sec:Perturbative-diagrammatics-deriv}
yields a non-iterative way to generate all non-standard diagrams.
We believe that this algorithm should be amenable to numerical implementation.
The diagrammatic Monte Carlo technique, for example, presents an effective
method to calculate the kernels of Schwinger-Dyson equations - for
the one-particle Greens function this is the self-energy \cite{Greitemann18,VanHoucke12_366,VanHoucke08_arxiv}.
One way to derive Schwinger-Dyson equations relies on multiple Legendre
transforms, expressing the generating functionals in terms of correlation
functions instead of potentials, as pioneered by de Dominicis and
Martin \cite{DeDominicis64_14}. The equations of state then constitute
self-consistency equations for these correlation functions, such as
eq. 6.11 in \cite{Vasiliev98}.  For example one needs to consider
the second Legendre transform to determine the self-energy and the
fourth to determine the two-particle Green's function self-consistently.
We do a first step towards developing diagrammatic rules for the second
Legendre transform in the appendix \prettyref{subsec:Second-Legendre-transform},
whereas higher order Legendre transforms are beyond the scope of the
current work. Their diagrammatics is already quite involved if the
underlying theory is Gaussian \cite[chap. 6.28.f]{Vasiliev98}; the
parquet equations form an approximation resulting from the fourth
order Legendre transform. A different way to obtain a set of self-consistency
equations, however, is by a direct resummation of diagrams, as in
the derivation of the Bethe-Salpeter equation \cite{Salpeter51}.
The diagrammatic rules derived here could be useful in such an approach.

In general, the methods also seems particularly promising for hierarchical
problems: assuming that a problem can be decomposed into small, but
strongly interacting clusters that can be solved exactly, the method
may be used to systematically expand in the interaction strength between
such clusters.

\section{Appendix}

\subsection{Definition of the effective action\label{app:Definition-effective-action}}

To define the effective action $\Gamma(x^{\ast})$ we eliminate the
dependence on the source field $j$ in favor of the mean value $x^{\ast}(j):=\langle x\rangle=\partial_{j}W(j)$
by using \prettyref{eq:def_partition} and \prettyref{eq:def_W}
and by following the usual background field method \cite[Chapter 3.23.6]{Kleinert09},
briefly summarized here: We express $W$ as the integral

\begin{eqnarray}
\exp\left(W(j)\right) & = & \Z(0)^{-1}\int_{x}\,\exp\left(S(x)+j^{\T}x\right).\label{eq:W_as_Z-1}
\end{eqnarray}
and then separate the fluctuations $\delta x=x-x^{\ast}$ from the
background value $x^{\ast}$ to get

\begin{eqnarray}
\exp\left(W(j)-j^{\T}x^{\ast}\right) & = & \Z(0)^{-1}\int_{\delta x}\,\exp\left(S(x^{\ast}+\delta x)+j^{\T}\delta x\right).\label{eq:pre_Legendre}
\end{eqnarray}
For given $x^{\ast}$, we now choose $j$ in a way that $x^{\ast}=\langle x\rangle(j)$
becomes the mean value of the field, so that $\delta x$ has vanishing
mean
\begin{eqnarray*}
0\stackrel{!}{=}\langle\delta x\rangle & \equiv & \Z(0)^{-1}\int_{\delta x}\,\exp\left(S(x^{\ast}+\delta x)+j^{\T}\delta x\right)\,\delta x\\
 & = & \Z(0)^{-1}\partial_{j}\,\int_{\delta x}\,\exp\left(S(x^{\ast}+\delta x)+j^{\T}\delta x\right)\\
 & = & \partial_{j}\,\exp\left(W(j)-j^{\T}x^{\ast}\right),
\end{eqnarray*}
where we used \prettyref{eq:pre_Legendre} in the last step. Since
the exponential function is monotonic, $\exp(x)^{\prime}>0\quad\forall x$,
the latter expression vanishes at the point where the exponent is
stationary
\begin{eqnarray}
\partial_{j}\left(W(j)-j^{\T}x^{\ast}\right) & = & 0,\label{eq:stationary_j-1}
\end{eqnarray}
The condition \prettyref{eq:stationary_j-1} has the form of a Legendre
transform \prettyref{eq:def_gamma} from the function $W(j)$ to the
effective action $\Gamma(x^{\ast})$. The supremum follows from stationarity
\prettyref{eq:stationary_j-1} and because $W$ is convex down, its
Hessian $W^{(2)}$, the covariance matrix, is positive definite (cf.
\prettyref{sub:Convexity-of-W} or \cite[p. 166]{ZinnJustin96}).
Therefore, $-W\left(j\right)+j^{T}x^{\ast}$, as a function of $j$,
is convex up (concave), so we may define the Legendre transform \prettyref{eq:def_gamma}
by the supremum over $j$.

\subsection{Convexity of $W$\label{sub:Convexity-of-W}}

$W$ is convex, because $W^{(2)}$ is the covariance matrix: it is
symmetric and therefore has real eigenvalues. For covariance matrices
these are in addition always positive \cite[p. 166]{ZinnJustin96}.
This can be seen from the following argument. Let us define the bilinear
form 
\begin{eqnarray*}
f(\eta) & : & =\eta^{\T}W^{(2)}\eta.
\end{eqnarray*}
A positive definite bilinear form has the property $f(\eta)>0\quad\forall\eta$.
With $\delta x:=x-\langle x\rangle$ we can express the covariance
as $W_{kl}^{(2)}=\langle\delta x_{k}\delta x_{l}\rangle$, so we may
explicitly write $f(\eta)$ as 
\begin{eqnarray*}
f(\eta) & = & \sum_{k,l}\eta_{k}W_{kl}^{(2)}\eta_{l}\\
 & = & \Z^{-1}(j)\,\eta^{\T}\int\,dx\,\delta x\,\delta x^{\T}\,\exp\left(S(x)+j^{\T}x\right)\eta\\
 & = & \Z^{-1}(j)\,\int\,dx\,\left(\eta^{\T}\delta x\right)^{2}\,\exp\left(S(x)+j^{\T}x)\right)>0,
\end{eqnarray*}
which is the expectation value of a positive quantity.

\subsection{Linked cluster theorem\label{app:Perturbative-cumulant-expansion}}

The following proof of connectedness of all diagrammatic contributions
to $W$ does not rely on the solvable part $S_{0}$ being Gaussian.
We here start from the general expression \prettyref{eq:def_partition}
to derive an expansion of $W(j)$, using the definition \prettyref{eq:def_W}
to write
\begin{eqnarray}
\exp(W(j))=Z(j) & = & \exp\left(\epsilon V(\partial_{j})\right)\,\exp\left(W_{0}(j)\right)\,\frac{\Z_{0}(0)}{\Z(0)}\label{eq:perturbation_general}
\end{eqnarray}

Taking the logarithm, the latter factor turns into an additive constant
$\ln\,\frac{\Z_{0}(0)}{\Z(0)}$ which ensures $W(0)=0$. Since we
are ultimately interested in the derivatives of $W$, namely the cumulants,
we may drop the constant and ensure $W(0)=0$ by finally dropping
the zeroth order Taylor coefficient.

The idea to prove connectedness follows to some extent \cite{ZinnJustin96}.
The proof is by induction, dissecting the operator $\exp\left(\epsilon V(\partial_{j})\right)$
appearing in \prettyref{eq:perturbation_general} into infinitesimal
operators using the definition of the exponential function as the
limit 
\begin{eqnarray}
\exp\left(\epsilon V(\partial_{j})\right) & = & \lim_{L\to\infty}(1+\frac{\epsilon}{L}\,V(\partial_{j}))^{L}.\label{eq:exponential_expansion}
\end{eqnarray}
For large $L$ given and fixed and $0\leq l\leq L$, we define
\begin{eqnarray*}
\exp\left(W_{l}(j)\right) & := & (1+\frac{\epsilon}{L}\,V(\partial_{j}))^{l}\,\exp\left(W_{0}(j)\right).
\end{eqnarray*}
It fulfills the trivial recursion $\exp(W_{l+1}(j))=(1+\frac{\epsilon}{L}\,V(\partial_{j}))\,W_{l}(j)$
from which follows an iteration by multiplication with $\exp(-W_{l}(j))$
and taking the logarithm
\begin{eqnarray}
 &  & W_{l+1}(j)-W_{l}(j)\label{eq:recursion_W_l}\\
 & = & \ln\,\left[\exp\left(-W_{l}(j)\right)\,(1+\frac{\epsilon}{L}\,V(\partial_{j}))\,\exp\left(W_{l}(j)\right)\right].\nonumber 
\end{eqnarray}
The desired result $W(j)$ then follows as the limit $W(j)=\lim_{L\to\infty}\,W_{L}(j).$
Expanding $\ln(1+\frac{\epsilon}{L}x)=\frac{\epsilon}{L}\,x+\mathcal{O}\left((\frac{\epsilon}{L})^{2}\right)$
in \prettyref{eq:recursion_W_l} we get
\begin{eqnarray}
 &  & W_{l+1}(j)-W_{l}(j)\label{eq:expansion_W_iteration}\\
 & = & \frac{\epsilon}{L}\,\left(\exp\left(-W_{l}(j)\right)\,V(\partial_{j})\,\exp\left(W_{l}(j)\right)\right)+O\big(\big(\frac{\epsilon}{L}\big)^{2}\big).\nonumber 
\end{eqnarray}
We start the induction by noting that for $l=0$ we have $W_{l=0}=W_{0}$,
the cumulant generating function of the solvable system. At this order,
$W$ hence does not contain any diagrammatic corrections; so in particular
no disconnected ones.

We assume that the assumption is true until some $0\le l\le L$. Stated
precisely, we assume that all perturbative corrections to $W_{l}(j)$
with $k$ vertices are connected and are $\propto\left(\frac{\epsilon}{L}\right)^{k}$;
the sub-leading terms $O\big(\big(\frac{\epsilon}{L}\big)^{2}\big)$
in \prettyref{eq:expansion_W_iteration} vanish in the limit $L\to\infty$,
as shown below. Representing the potential $V$ as a Taylor series,
we see that each step adds terms of the form shown in the second
line
\begin{eqnarray}
W_{l+1}(j) & = & 1\cdot W_{l}(j)\label{eq:iterative_W}\\
 & + & \frac{\epsilon}{L}\cdot\sum_{\{n_{i}\}}\frac{V^{(n_{1},\ldots,n_{N})}}{n_{1}!\cdots n_{N}!}\,\exp\left(-W_{l}(j)\right)\,\partial_{1}^{n_{1}}\cdots\partial_{N}^{n_{N}}\,\exp\left(W_{l}(j)\right),\nonumber 
\end{eqnarray}
where $\partial_{i}$ is used in short for $\partial_{j_{i}}$. The
Taylor coefficient $\frac{V^{(n_{1},\ldots,n_{N})}}{n_{1}!\cdots n_{N}!}$
is graphically represented by a vertex (see \prettyref{fig:Expansion-example}).
Noting that the two exponential factors cancel each other after the
differential operator has been applied to the latter factor, what
remains is a set of connected components of $W_{l}(j)$ tied together
by the vertex $\frac{V^{(n_{1},\ldots,n_{N})}}{n_{1}!\cdots n_{N}!}$.
Disconnected components cannot appear, because there is only a single
vertex; each of its legs belongs to one $\partial_{i}$, which, by
acting on $W_{l}(j)$, attaches to one leg of the components in $W_{l}$.

The iteration \prettyref{eq:iterative_W} shows that the second term
in each step adds to $W_{l}(j)$ a set of diagrams to obtain $W_{l+1}(j)$.
It is clear from the single appearance of $V^{(n_{1},\ldots,n_{N})}$
that in each iteration only one such additional vertex is added. We
will show now that we, moreover, only need to consider such additional
diagrams, where the new vertex $V^{(n_{1},\ldots,n_{N})}$ connects
to a perturbative correction contained in $W_{l}$ (with $k\ge1$
vertices), while all of its remaining legs connect to a cumulant of
the unperturbed theory $W_{0}$ (with $k=0$ vertices). Stated differently,
a perturbative correction with $k$ vertices picks up each of its
vertices in a different iteration step $l$ in \prettyref{eq:iterative_W};
contributions where a single iteration step increases the number of
vertices in a component by more than one vanish in the $L\to\infty$
limit.

To understand why this is so, we consider the overall factor in front
of a resulting diagram with $k$ vertices after $L$ iterations of
\prettyref{eq:iterative_W}. In each step of \prettyref{eq:iterative_W}
the first term copies all diagrams from $W_{l}$. The second term
adds those formed by help of the additional vertex $V^{(n_{1},\ldots,n_{N})}$.
Following how one particular graph is generated by the iteration,
in each step we have the binary choice to either leave it as it is
or to combine it with other components by help of an additional vertex.

We first consider the case that each of the $k$ vertices is picked
up in a different step (at different $l$) in the iteration. Each
such step comes with a factor $\frac{\epsilon}{L}$ and in there are
$\left(\begin{array}{c}
L\\
k
\end{array}\right)$ ways to select $k$ out of the $L$ iteration steps to pick up a
vertex to construct this particular diagram. So in total we get a
factor
\begin{eqnarray}
\left(\frac{\epsilon}{L}\right)^{k}\left(\begin{array}{c}
L\\
k
\end{array}\right) & = & \frac{\epsilon^{k}}{k!}\,\frac{L(L-1)\cdots(L-k+1)}{L^{k}}\;\stackrel{L\to\infty}{\to}\,\frac{\epsilon^{k}}{k!},\label{eq:factor_pick_up_one}
\end{eqnarray}
which is independent of $L$.

Now consider the case that we pick up the $k$ vertices along the
iteration \prettyref{eq:iterative_W} such that in one step we combined
two sub-components with each one or more vertices already. Consequently,
to arrive at $k$ vertices in the end, we only need $k^{\prime}<k$
iteration steps in which the second rather than the first term of
\prettyref{eq:iterative_W} acted on the component. The overall factor
therefore is
\begin{eqnarray}
\left(\frac{\epsilon}{L}\right)^{k}\left(\begin{array}{c}
L\\
k^{\prime}
\end{array}\right) & = & \frac{\epsilon^{k}}{k^{\prime}!}\,\frac{L(L-1)\cdots(L-k^{\prime}+1)}{L^{k}}\label{eq:factor_pick_up_many}\\
 & \stackrel{L\gg k^{\prime}}{=} & \frac{\epsilon^{k}}{k^{\prime}!}\,\frac{1}{L^{k-k^{\prime}}}\stackrel{L\to\infty}{=}0.\nonumber 
\end{eqnarray}
We can hence neglect the latter option and conclude that $W(j)$ is
composed of all connected graphs, where a perturbative correction
with $k$ vertices comes with the factor $\frac{\epsilon^{k}}{k!}$
given by \prettyref{eq:factor_pick_up_one}. By the same reasoning
we may neglect the $O\big(\big(\frac{\epsilon}{L}\big)^{2}\big)$
term in \prettyref{eq:expansion_W_iteration}, because also here in
a single step we would increase the order of the diagram by more than
one factor $L^{-1}$.

\subsection{Operator equation for $\Gamma_{V}$\label{app:operator_equation_Gamma_V}}

Let us first see why the decomposition into a sum in \prettyref{eq:Gamma_pert_decomposition}
holds. To this end, we consider \prettyref{eq:int_diff_gamma} with
$\delta x=x-x^{\ast}$ and use the decomposition \prettyref{eq:def_S_pert}
of the action as well as the decomposition \prettyref{eq:Gamma_pert_decomposition}
of $\Gamma$ to obtain
\begin{eqnarray*}
 &  & \exp(-\Gamma_{0}(x^{\ast})-\Gamma_{V}(x^{\ast}))\\
 & = & \Z^{-1}(0)\,\int_{x}\,\exp\big(S_{0}(x)+\epsilon V(x)+\big(\Gamma_{0}^{(1)\T}(x^{\ast})+\Gamma_{V}^{(1)\T}(x^{\ast})\big)(x-x^{\ast})\big).
\end{eqnarray*}
Collecting the terms depending on $x^{\ast}$ on the left hand side
we get with \prettyref{eq:equn_of_state}

\begin{eqnarray*}
 &  & \exp(\underbrace{-\Gamma_{0}(x^{\ast})+\Gamma_{0}^{(1)\T}(x^{\ast})\,x^{\ast}}_{W_{0}(j)\big|_{j=\Gamma_{0}^{(1)}(x^{\ast})}}-\Gamma_{V}(x^{\ast}))\\
 & = & \exp\big(\epsilon V(\partial_{j})+\Gamma_{V}^{(1)\T}(x^{\ast})(\partial_{j}-x^{\ast})\big)\,\Z^{-1}(0)\int_{x}\,\exp\big(S_{0}(x)+j^{\T}x)\big)\big|_{j=\Gamma_{0}^{(1)}(x^{\ast})}\\
 & = & \exp\big(\epsilon V(\partial_{j})+\Gamma_{V}^{(1)\T}(x^{\ast})(\partial_{j}-x^{\ast})\big)\,\exp\big(W_{0}(j)\big)\big|_{j=\Gamma_{0}^{(1)}(x^{\ast})},
\end{eqnarray*}
where we moved the perturbing potential in front of the integral,
making the replacement $x\to\partial_{j}$ and we identified the unperturbed
cumulant generating function $\exp(W_{0}(j))=\Z^{-1}(0)\,\int_{x}\,\exp\big(S_{0}(x)+j^{\T}x\big)$
from the second to the third line. With the term $\Gamma_{0}^{(1)\T}(x^{\ast})x^{\ast}\equiv j_{0}^{\T}x^{\ast}$
on the left hand side, we get $-\Gamma_{0}(x^{\ast})+j^{\T}x^{\ast}=W_{0}(j)\big|_{j=\Gamma_{0}^{(1)}(x^{\ast})}$,
which follows from the definition \prettyref{eq:Gamma0_pert-1}. Multiplying
with $\exp(-W_{0}(j))\big|_{j=\Gamma_{0}^{(1)}(x^{\ast})}$ from left
then leads to a recursive equation \prettyref{eq:Gamma_V_recursion}
for $\Gamma_{V}$, which shows that our ansatz \prettyref{eq:Gamma_pert_decomposition}
was indeed justified and that we may determine $\Gamma_{V}$ recursively,
since $\Gamma_{V}$ appears again on the right hand side.

\subsection{Recursion for $g_{l}$\label{app:Recusion-equation-for-Gamma_V}}

To construct the diagrams iteratively we write the perturbing term
in \prettyref{eq:Gamma_V_recursion} as 
\begin{eqnarray}
 &  & \exp\big(\epsilon V(\partial_{j})+\Gamma_{V}^{(1)\T}(x^{\ast})(\partial_{j}-x^{\ast})\big)\label{eq:pert_term_factors}\\
 & = & \lim_{L\to\infty}\left(1+\frac{1}{L}\left(\epsilon V(\partial_{j})+\Gamma_{V}^{(1)\T}(x^{\ast})(\partial_{j}-x^{\ast})\right)\right)^{L}.\nonumber 
\end{eqnarray}
Inserted into \prettyref{eq:Gamma_V_recursion} we assume $L$ fixed
but large and choose some $0\le l\le L$. We define $G_{l}(j)$ as
the result after application of $l$ factors of the right hand side
of \prettyref{eq:pert_term_factors}
\begin{eqnarray*}
\exp(G_{l}(j)) & := & \left(1+\frac{1}{L}\left(\epsilon V(\partial_{j})+\Gamma_{V}^{(1)\T}(x^{\ast})\,(\partial_{j}-x^{\ast})\right)\right)^{l}\,\exp(W_{0}(j)).
\end{eqnarray*}
Obviously we have 
\begin{eqnarray}
G_{0} & \equiv & W_{0}.\label{eq:G_0_initial}
\end{eqnarray}
For $l=L\to\infty$, we obtain the desired result as 
\begin{eqnarray*}
-\Gamma_{V}(x^{\ast}) & = & \lim_{L\to\infty}\,G_{L}(j)-W_{0}(j)\Big|_{_{j=\Gamma_{0}^{(1)}(x^{\ast})}}.
\end{eqnarray*}
By its definition, $G_{l}(j)$ obeys the trivial iteration 
\begin{eqnarray*}
 &  & \exp(G_{l+1}(j))\\
 & = & \left(1+\frac{1}{L}\left(\epsilon V(\partial_{j})+\Gamma_{V}^{(1)\T}(x^{\ast})\,(\partial_{j}-x^{\ast})\right)\right)\,\exp(G_{l}(j)).
\end{eqnarray*}
Multiplying from left with $\exp(-G_{l}(j))$ and taking the logarithm
on both sides while using $\ln(1+\frac{1}{L}x)=\frac{1}{L}x+O(L^{-2})$
we get the recursion for the additional diagrams produced in step
$l+1$
\begin{eqnarray}
 &  & G_{l+1}(j)-G_{l}(j)\label{eq:iteration_G_app}\\
 & = & \frac{\epsilon}{L}\,\exp(-G_{l}(j))\,V(\partial_{j})\,\exp(G_{l}(j))\nonumber \\
 & + & \frac{1}{L}\,\exp(-G_{l}(j))\,\left(\Gamma_{V}^{(1)\T}(x^{\ast})\,(\partial_{j}-x^{\ast})\right)\,\exp(G_{l}(j))\nonumber \\
 & + & \mathcal{O}(L^{-2}).\nonumber 
\end{eqnarray}
By the initial condition \prettyref{eq:G_0_initial} and the form
of the additive iteration \prettyref{eq:iteration_G_app} it is clear
that all graphs of $W_{0}$ are also contained in $G_{l}$ for any
step $l$. We may therefore define only the perturbative corrections
as $g_{l}:=G_{l}-W_{0}$.

We first note that indeed \prettyref{eq:iteration_G_app} yields a
closed iteration: Constructing the graphs of up to order $l+1$ in
$G_{l+1}$ by \prettyref{eq:iteration_G_app} we only need the graphs
in $G_{l}$ on the right hand side of \prettyref{eq:iteration_G_app},
which, by construction, are of order $\le l$ . This is because $V$
contains exactly one bare vertex and $\Gamma_{V}^{(1)}(x^{\ast})$
contains at least one.

 Taken together, we arrive at the central result of our work, Eq.
\prettyref{eq:iteration_g}.

\subsection{Second Legendre transform\label{subsec:Second-Legendre-transform}}

We here extend the iterative procedure to compute perturbative corrections
to the second Legendre transform. To this end, we define an effective
action that is a function of the first moment $x^{\ast}$ and the
second moment $c^{\ast}=\langle x^{2}\rangle$. We express $W$ as
the integral

\begin{eqnarray}
\exp\left(W(j,k)\right) & = & \Z(0)^{-1}\int_{x}\,\exp\left(S(x)+j^{\T}x+k^{\T}x^{2}\right).\label{eq:W_as_Z-1-1}
\end{eqnarray}
Here $k^{\T}x^{2}$ must be understood as a bilinear form in $x$,
hence $\sum_{il}k_{il}x_{i}x_{l}$. We have
\begin{eqnarray*}
\frac{\partial W}{\partial j} & = & \langle x\rangle=:x^{\ast},\qquad\frac{\partial W}{\partial k}=\langle x^{2}\rangle=:c^{\ast}.
\end{eqnarray*}
We would like to define an effective action that is a function of
these latter coordinates. So we define $\Gamma(x^{\ast},c^{\ast})$
as

\begin{eqnarray}
\exp\left(-\Gamma(x^{\ast},c^{\ast})\right) & = & \Z(0)^{-1}\int_{x}\,\exp\left(S(x)+j^{\T}(x-x^{\ast})+k^{\T}(x^{2}-c^{\ast})\right)\nonumber \\
\label{eq:pre_legendre-1}\\
\Gamma(x^{\ast},c^{\ast}) & = & j^{\T}x^{\ast}+k^{\T}c^{\ast}-W(j,k),\nonumber 
\end{eqnarray}
with the additional constraints that $\partial/\partial j$ and $\partial/\partial k$
of the right hand side vanishes. Consequently, the so-defined function
$\Gamma$ fulfills the two equations of state
\begin{eqnarray*}
\frac{\partial\Gamma}{\partial x^{\ast}} & = & j,\qquad\frac{\partial\Gamma}{\partial c^{\ast}}=k,
\end{eqnarray*}
which can be obtained by application of the chain rule. The second
of these equations, for example, follows as
\begin{eqnarray*}
\frac{\partial\Gamma(x^{\ast},c^{\ast})}{\partial c^{\ast}} & = & \frac{\partial j^{\T}}{\partial c^{\ast}}x^{\ast}+k+\frac{\partial k^{\T}}{\partial c^{\ast}}c^{\ast}-\underbrace{\frac{\partial W^{\T}}{\partial j}}_{\equiv x^{\ast}}\frac{\partial j}{\partial c^{\ast}}-\underbrace{\frac{\partial W^{\T}}{\partial k}}_{\equiv c^{\ast}}\frac{\partial k}{\partial c^{\ast}}=k.
\end{eqnarray*}
 Hence we may write the definition of $\Gamma$ also as
\begin{eqnarray*}
\exp\left(-\Gamma(x^{\ast},c^{\ast})\right) & = & \Z(0)^{-1}\int_{x}\,\exp\left(S(x)+\frac{\partial\Gamma^{\T}}{\partial x^{\ast}}\,(x-x^{\ast})+\frac{\partial\Gamma^{\T}}{\partial c^{\ast}}\,(x^{2}-c^{\ast})\right).
\end{eqnarray*}
We next decompose $S(x)=S_{0}(x)+\epsilon V(x)$, where we assume
that we may compute the cumulant generating function $W_{0}(j,k)=\ln\int_{x}\,\exp(S_{0}(x)+j^{\T}x+k^{\T}x^{2})$
exactly. Correspondingly, we assume a decomposition of the effective
action into the solvable part $\Gamma_{0}$ and the perturbative corrections
$\Gamma_{V}$ as
\begin{eqnarray*}
\Gamma(x^{\ast},c^{\ast}) & = & \Gamma_{0}(x^{\ast},c^{\ast})+\Gamma_{V}(x^{\ast},c^{\ast}).
\end{eqnarray*}
With the notation $\frac{\partial\Gamma}{\partial x^{\ast}}=:\Gamma^{(1,0)}$
and $\frac{\partial\Gamma}{\partial c^{\ast}}=:\Gamma^{(0,1)}$ we
may express the integral equation \prettyref{eq:pre_legendre-1} as

\begin{eqnarray*}
 &  & \exp(-\Gamma_{0}(x^{\ast},c^{\ast})-\Gamma_{V}(x^{\ast},c^{\ast}))\\
 & = & \Z^{-1}(0)\,\int_{x}\,\exp\Big(S_{0}(x)+\epsilon V(x)\\
 &  & \phantom{\Z^{-1}(0)\,\int_{x}\,\exp}+\big(\Gamma_{0}^{(1,0)\T}+\Gamma_{V}^{(1,0)}\big)^{\T}(x-x^{\ast})\\
 &  & \phantom{\Z^{-1}(0)\,\int_{x}\,\exp}+\big(\Gamma_{0}^{(0,1)\T}+\Gamma_{V}^{(0,1)}\big)^{\T}(x^{2}-c^{\ast})\Big)
\end{eqnarray*}
or, bringing all terms that are independent of the integration variable
$x$ to the left, we get with $j_{0}=\Gamma_{0}^{(1,0)}$ and $k_{0}=\Gamma_{0}^{(0,1)}$
\begin{eqnarray*}
 &  & \exp(\underbrace{-\Gamma_{0}(x^{\ast},c^{\ast})+j_{0}^{\T}x^{\ast}+k_{0}^{\T}c^{\ast}}_{W_{0}(j_{0},k_{0})}-\Gamma_{V}(x^{\ast},c^{\ast}))\\
 & = & \exp\big(\epsilon V(\partial_{j})+\Gamma_{V}^{(1,0)\T}(\partial_{j}-x^{\ast})+\Gamma_{V}^{(0,1)\T}\,(\partial_{k}-c^{\ast})\big)\times\\
 &  & \times\Z^{-1}(0)\int_{x}\,\exp\big(S_{0}(x)+j^{\T}x+k^{\T}x^{2}\big)\big|_{j=\Gamma_{0}^{(1,0)}(x^{\ast},c^{\ast}),\,k=\Gamma_{0}^{(0,1)}(x^{\ast},c^{\ast}).}
\end{eqnarray*}
We hence obtain the operator form of the equation as
\begin{eqnarray*}
\exp(-\Gamma_{V}(x^{\ast},c^{\ast})) & = & \exp(-W_{0}(j,k))\times\\
 & \times & \exp\big(\epsilon V(\partial_{j})+\Gamma_{V}^{(1,0)\T}(\partial_{j}-x^{\ast})+\Gamma_{V}^{(0,1)\T}(\partial_{k}-c^{\ast})\big)\times\\
 & \times & \exp(W_{0}(j,k))\big|_{j=\Gamma_{0}^{(1,0)}(x^{\ast},c^{\ast}),k=\Gamma_{0}^{(0,1)}(x^{\ast},c^{\ast})}.
\end{eqnarray*}
Using the representation of the exponential function as a series and
the series expansion of the logarithm to first order, we get the iteration

\begin{eqnarray}
 &  & g_{l+1}(j,k)-g_{l}(j,k)\label{eq:iteration_g-1}\\
 & = & \frac{\epsilon}{L}\,\exp(-W_{0}(j,k)-g_{l}(j,k))\,V(\partial_{j})\,\exp(W_{0}(j,k)+g_{l}(j,k))\nonumber \\
 & + & \frac{1}{L}\,\exp(-W_{0}(j,k)-g_{l}(j,k))\,\Gamma_{V}^{(1,0)}\left(\partial_{j}-x^{\ast}\right)\,\exp(W_{0}(j,k)+g_{l}(j,k))\nonumber \\
 & + & \frac{1}{L}\,\exp(-W_{0}(j,k)-g_{l}(j,k))\,\Gamma_{V}^{(0,1)}\left(\partial_{k}-c^{\ast}\right)\,\exp(W_{0}(j,k)+g_{l}(j,k))\nonumber \\
 & + & \mathcal{O}(L^{-2}),\nonumber 
\end{eqnarray}
with the initial condition $g_{0}=0$. The perturbative corrections
to the effective action then result as the limit
\begin{eqnarray*}
-\Gamma_{V}(x^{\ast},c^{\ast})= & \lim_{L\to\infty} & \,g_{L}(j,k)\Big|_{j=\Gamma_{0}^{(1,0)}(x^{\ast},c^{\ast}),\,k=\Gamma_{0}^{(0,1)}(x^{\ast},c^{\ast})}.
\end{eqnarray*}
One could use this iterative procedure to compute perturbative corrections.
It seems that a proof of a generalized irreducibility should follow
along similar lines as in the case of a first order Legendre transform.
In particular, the term $\Gamma_{V}^{(0,1)}\left(\partial_{k}-c^{\ast}\right)$
establishes a double link between a component contained in $g_{l}$
and one in $\Gamma_{V}$, thus suggesting a more general 2PI property.
We will leave a more careful consideration open for subsequent works.

\subsection{Taxonomy of reducible diagrams\label{subsec:Taxonomy-of-reducible}}

One might wonder why the diagram in \prettyref{fig:reducible-diagrams-generated}c)
remains in the perturbative expansion of $\Gamma$ despite being reducible
in the sense that it is divided into two parts by cutting one of the
legs of $\Gamma_{0}^{\left(2\right)}$, which otherwise acts similarly
as a leg of a bare interaction vertex. We here provide an explanation
in the framework of the iterative construction of $\Gamma_{V}$. There
are four different types of diagrams that are reducible in the sense
that they fall apart if one detaches one leg from a cumulant. We
classify them by the type of subdiagrams (leaves) they are composed
of and how these leaves are connected. We can distinguish five cases:
\begin{enumerate}
\item[\foreignlanguage{english}{I}] Two irreducible diagrams from $W_{V}$ connected by a single leg
that belongs to a bare interaction vertex.\label{enu:I-Ttwo--diagrams-Wv}
\item[\foreignlanguage{english}{Ia}]  Multiple irreducible diagrams from $W_{V}$ that are connected to
the remainder by the same junction as in case I.\label{enu:-Multiple-diagrams-Wv}
\item[\foreignlanguage{english}{II}]  Two irreducible diagrams from $W_{V}^{\left(1\right)}$ connected
via a $\Gamma_{0}^{\left(2\right)}$-component.\label{enu:-Two-diagramsWv-Gamma0}
\item[\foreignlanguage{english}{III}]  One diagram from $W_{V}^{\left(1\right)}$ and one non-standard
component of $\Gamma_{V}$ (not contained in the expansion of $W_{V}$),
necessarily connected by a $\Gamma_{0}^{\left(n\right)}$-component\label{enu:-One-diagram-Wv_One_Gammav}
\item[\foreignlanguage{english}{IV}]  Two non-standard diagrams of $\Gamma_{V}$ (not contained in the
expansion of $W_{V}$), necessarily connected by a $\Gamma_{0}^{\left(n\right)}$-component\label{enu:-Two-components_Gammav}
\end{enumerate}
In case I, in which both parts are irreducible, we may insert either
one or two unities of the form
\begin{equation}
1=\Diagram{gpgfcf}
,\ \label{eq:Decomposition_of_Unity}
\end{equation}
where empty and hatched circles in this section always denote the
unperturbed quantities $W_{0}^{\left(n\right)}$ and $\Gamma_{0}^{\left(2\right)}$,
respectively. We therefore see that we can replace both leaves by
a part of $\Gamma_{V}^{\left(1\right)}\left(x^{\ast}\right)$. So,
together with the original diagram, we have four contributions that
all contribute with the same magnitude, two of them positive, two
of them negative: The sum is $0$. Schematically, the situation is
as follows:

\begin{fmffile}{Example_cancelling}	
	\begin{eqnarray}
	\mkern-108mu 0=\quad\parbox{75mm}{
		\begin{fmfgraph*}(70,30)
			\fmfpen{0.75thin}
			\fmfleft{i1}
			\fmfright{o1}
			\fmf{plain, tension=.25, right=.5}{i1,v3,i1}
			\fmf{plain, tension=.25, right=.25}{i1,v3,i1}
			\fmf{plain, tension=1.}{v3,v4}
			\fmf{plain,tension=.33, right=.5}{v4,o1,v4}
			\fmf{plain,tension=.33}{v4,o1}
			\fmfv{decor.shape=circle, decor.filled=empty, decor.size = thin}{v3}
			\fmfv{d.s=circle, d.filled=hatched}{i1,o1}
		\end{fmfgraph*}		
	}
	\mkern-236mu
	& -2 \quad
	& \mkern-18mu  \quad \parbox{125mm}{
		\begin{fmfgraph*}(120,30)
			\fmfpen{0.75thin}
			\fmfleft{i1}
			\fmfright{o1}
			\fmf{plain, tension=.25, right=.5}{i1,v3,i1}
			\fmf{plain, tension=.25, right=.25}{i1,v3,i1}
			\fmf{plain, tension=1.}{v3,g1}
			\fmf{plain, tension=1.}{g1,c1}
			\fmf{plain, tension=1.5}{c1,v4}
			\fmf{plain,tension=.33, right=.5}{v4,o1,v4}
			\fmf{plain,tension=.33}{v4,o1}
			\fmfv{decor.shape=circle, decor.filled=empty, decor.size = thin}{v3,c1}
			\fmfv{decor.shape=circle, decor.filled=shaded, decor.size = thin}{g1}
			\fmfv{d.s=circle, d.filled=hatched}{i1,o1}
			\fmfv{label=$\underbrace{\phantom{ist Phantomas!}}_{\subseteq \Gamma^{\left(1\right)}_{V}}$, label.angle=-90, label.dist=5pt}{v3}
			\fmfv{label=$\overbrace{\phantom{ente das ist Phantomas!}}^{\subseteq \Gamma^{\left(1\right)}_{V}}$, label.angle=110, label.dist=5pt}{v4}
		\end{fmfgraph*}		
	} \nonumber\\
	& & \nonumber\\
	&\mkern-18mu +&
	\parbox{125mm}{
		\begin{fmfgraph*}(180,30)
			\fmfpen{0.75thin}
			\fmfleft{i1}
			\fmfright{o1}
			\fmf{plain, tension=.25, right=.5}{i1,v3,i1}
			\fmf{plain, tension=.25, right=.25}{i1,v3,i1}
			\fmf{plain, tension=1.}{v3,g1}
			\fmf{plain, tension=1.}{g1,c1}
			\fmf{plain, tension=1.}{c1,g2}
			\fmf{plain, tension=1.}{g2,c2}
			\fmf{plain, tension=1.5}{c2,v4}
			\fmf{plain,tension=.33, right=.5}{v4,o1,v4}
			\fmf{plain,tension=.33}{v4,o1}
			\fmfv{decor.shape=circle, decor.filled=empty, decor.size = thin}{v3,c1,c2}
			\fmfv{decor.shape=circle, decor.filled=shaded, decor.size = thin}{g1,g2}
			\fmfv{d.s=circle, d.filled=hatched}{i1,o1}
			\fmfv{label=$\underbrace{\phantom{ist Phantomas!}}_{\subseteq \Gamma^{\left(1\right)}_{V}}$, label.angle=-90, label.dist=5pt}{v3}
			\fmfv{label=$\overbrace{\phantom{ente das ist Phantomas!}}^{\subseteq \Gamma^{\left(1\right)}_{V}}$, label.angle=110, label.dist=5pt}{v4}
		\end{fmfgraph*}		
	}\mkern-268mu.\nonumber\\ \nonumber
	\end{eqnarray} 
\end{fmffile}

Here the circles filled with squares denote some arbitrary continuation
of the respective diagram. We conclude that in second order in the
interactions, no diagrams with irreducible $W_{V}^{\left(1\right)}$-leaves
contribute. We will proof by induction in the number of vertices $n$
that no diagram described in case Ia (with at least one reducible
link) contributes to $\Gamma_{V}$. Case I is our induction assumption
($n=2$). Assume that the assumption holds $\forall\,k\leq n$. We
observe that if the whole diagram contains $n+1$ interactions we
can pick an arbitrary leaf that is connected to the rest by a single
link, so that the (reducible) rest contains at most $n$ interactions.
By the induction assumption, this rest is not contained in the diagrammatic
expansion of $\Gamma_{V}$. Therefore, we have just one contribution,
which can be replaced by a part of $\Gamma_{V}^{\left(1\right)}$
contributing with opposite sign. This contribution cancels the contribution
constructed solely out of bare vertices. Diagrammatically this could
be depicted by\begin{fmffile}{Example_cancelling_single}	
	\begin{eqnarray}
	\mkern-36mu 0=\quad\parbox{75mm}{
		\begin{fmfgraph*}(70,30)
			\fmfpen{0.75thin}
			\fmfleft{i1}
			\fmfright{o1}
			\fmf{plain, tension=.25, right=.5}{i1,v3,i1}
			\fmf{plain, tension=.25, right=.25}{i1,v3,i1}
			\fmf{plain, tension=1.}{v3,v4}
			\fmf{plain,tension=.33, right=.5}{v4,o1,v4}
			\fmf{plain,tension=.33}{v4,o1}
			\fmfv{decor.shape=circle, decor.filled=empty, decor.size = thin}{v3}
			\fmfv{d.s=circle, d.filled=hatched}{i1,o1}
		\end{fmfgraph*}		
	}
	\mkern-236mu
	& - \quad
	& \mkern-18mu  \quad \parbox{125mm}{
		\begin{fmfgraph*}(120,30)
			\fmfpen{0.75thin}
			\fmfleft{i1}
			\fmfright{o1}
			\fmf{plain, tension=.25, right=.5}{i1,v3,i1}
			\fmf{plain, tension=.25, right=.25}{i1,v3,i1}
			\fmf{plain, tension=1.}{v3,g1}
			\fmf{plain, tension=1.}{g1,c1}
			\fmf{plain, tension=1.5}{c1,v4}
			\fmf{plain,tension=.33, right=.5}{v4,o1,v4}
			\fmf{plain,tension=.33}{v4,o1}
			\fmfv{decor.shape=circle, decor.filled=empty, decor.size = thin}{v3,c1}
			\fmfv{decor.shape=circle, decor.filled=shaded, decor.size = thin}{g1}
			\fmfv{d.s=circle, d.filled=hatched}{i1,o1}
			\fmfv{label=$\underbrace{\phantom{ist Phantomas!}}_{\subseteq \Gamma^{\left(1\right)}_{V}}$, label.angle=-90, label.dist=5pt}{v3}
		\end{fmfgraph*}		
	}\mkern-400mu. \nonumber \\ \nonumber
	\end{eqnarray} 
\end{fmffile}

This concludes the induction. We conclude from these observations
that no diagrams with a single irreducible $W_{V}^{\left(1\right)}$-leaf
contribute. 

Considering case II, it is obvious that we can choose one of the leaves
to be provided by $\Gamma_{V}^{\left(1\right)}$ and the other one
to be composed of bare vertices. We can also insert a unity \prettyref{eq:Decomposition_of_Unity}
to the left of the given $\Gamma_{0}^{\left(2\right)}$, so that we
can interpret both leaves to come from $\Gamma_{V}^{\left(1\right)}\left(x^{\ast}\right)$.
Because the diagram composed solely of bare vertices is missing, we
only have three contributions. They all contribute with the same absolute
value, but two of them with one sign, one with the other sign and
therefore, the contributions do not sum to zero and the whole diagram
contributes to the expansion of $\Gamma_{V}$, as claimed in the main
text. Diagrammatically, this is illustrated by

\begin{fmffile}{Example_non_cancelling}	
	\begin{eqnarray}
	\mkern-90mu 0\neq-2\cdot \quad \parbox{125mm}{
		\begin{fmfgraph*}(105,30)
			\fmfpen{0.75thin}
			\fmfleft{i1}
			\fmfright{o1}
			\fmf{plain, tension=.25, right=.5}{i1,v3,i1}
			\fmf{plain, tension=.25, right=.25}{i1,v3,i1}
			\fmf{plain, tension=1.}{v3,g1}
			\fmf{plain, tension=1.}{g1,v4}
			\fmf{plain,tension=.33, right=.5}{v4,o1,v4}
			\fmf{plain,tension=.33}{v4,o1}
			\fmfv{decor.shape=circle, decor.filled=empty, decor.size = thin}{v3,v4}
			\fmfv{decor.shape=circle, decor.filled=shaded, decor.size = thin}{g1}
			\fmfv{d.s=circle, d.filled=hatched}{i1,o1}
			\fmfv{label=$\underbrace{\phantom{ist Phantomas!}}_{\subseteq \Gamma^{\left(1\right)}_{V}}$, label.angle=-90, label.dist=5pt}{v3}
			\fmfv{label=$\overbrace{\phantom{ist Phantomas!}}^{\subseteq \Gamma^{\left(1\right)}_{V}}$, label.angle=110, label.dist=5pt}{v4}
		\end{fmfgraph*}		
	}
	\mkern-420mu
	& + \quad
	& 
	\parbox{125mm}{
		\begin{fmfgraph*}(150,30)
			\fmfpen{0.75thin}
			\fmfleft{i1}
			\fmfright{o1}
			\fmf{plain, tension=.25, right=.5}{i1,v3,i1}
			\fmf{plain, tension=.25, right=.25}{i1,v3,i1}
			\fmf{plain, tension=1.}{v3,g1}
			\fmf{plain, tension=1.}{g1,c1}
			\fmf{plain, tension=1.}{c1,g2}
			\fmf{plain, tension=1.}{g2,v4}
			\fmf{plain,tension=.33, right=.5}{v4,o1,v4}
			\fmf{plain,tension=.33}{v4,o1}
			\fmfv{decor.shape=circle, decor.filled=empty, decor.size = thin}{v3,c1,v4}
			\fmfv{decor.shape=circle, decor.filled=shaded, decor.size = thin}{g1,g2}
			\fmfv{d.s=circle, d.filled=hatched}{i1,o1}
			\fmfv{label=$\underbrace{\phantom{ist Phantomas!}}_{\subseteq \Gamma^{\left(1\right)}_{V}}$, label.angle=-90, label.dist=5pt}{v3}
			\fmfv{label=$\overbrace{\phantom{ist Phantomas!}}^{\subseteq \Gamma^{\left(1\right)}_{V}}$, label.angle=110, label.dist=5pt}{v4}
		\end{fmfgraph*}		
	}\mkern-350mu. \nonumber\\
	\end{eqnarray} 
\end{fmffile}

In point II, we have proved that there are diagrams contributing to
$\Gamma_{V}$, but not to $W_{V}$ (for the example \prettyref{fig:reducible-diagrams-generated}c)),
therefore the situations of points III and IV can appear.

The diagrams described under point III cancel, because the irreducible
components within the $W_{V}^{\left(1\right)}$-like leaf can be replaced
a $\Gamma_{V}^{\left(1\right)}$-part or can be left as it is. Both
contribute with same absolute value, but opposite sign and cancel.
Because the other part cannot be composed just by bare vertices and
unperturbed cumulants, it occurs only once, namely as part of $\Gamma_{V}^{\left(1\right)}$.
Therefore, the contribution from this diagram remains zero.

Point IV is even simpler, because by construction we only have a unique
way to construct the composed diagram, namely by joining two $\Gamma_{V}^{\left(1\right)}$-parts
and therefore, the whole diagram is not canceled.

Using these rules to determine if a given diagram contributes to $\Gamma_{V}$
or not, we conclude from case Ia that, as mentioned in the main text,
diagrams with at least one component from $W_{V}$ do not contribute.
However, one has to be careful applying case III to rule out the occurrence
of a certain diagram, that can be decomposed according to III, because
this decomposition is not necessarily unique and so the contribution
of the whole diagram is actually not zero. 

\subsection{Calculation of symmetry factors in the diagrammatic solution of the
Ising model\label{app:App_Calc_of_sym_factors}}

We determine here the symmetry factors of the diagrams given in the
main text using the following scheme: First, label all legs of interactions
by indices. Then, count the possible combinations of swaps of the
two legs at an interaction and permutations of interactions that lead
to a new labeled diagram. 

For the second-order-diagram (\ref{Def_second_oder_diagram}), this
is just $2$ coming from the fact that flipping a single vertex produces
a new labeled diagram, but flipping both vertices leaves the diagram
invariant.

For the ring diagram in third order (\ref{Def_third_oder_ring_diagram}),
permuting interactions produces the same diagram (``at most'' it
mirrors the diagram). The same holds for the other third order diagram
(\ref{Def_third_oder_Citroen_diagram}).

Swapping legs in the last diagram (\ref{Def_third_oder_Citroen_diagram})
yields a factor $2^{2}$, because switching all vertex legs does not
yield a new labeled diagram, whereas for the ring diagram, it does,
leading to the symmetry factor $2^{3}$.

Proceeding to fourth order, we first compute the symmetry factor $\SR$
of the ring diagram in (\ref{Fourth_order_ring}). For a given interaction,
we have $3$ possibilities to choose another interaction \textit{not
}to pair it with and every vertex flip produces a new labeled diagram.
This gives $S_{\mathrm{R}}=3\cdot2^{4}=48.$ 

For the second diagram (\ref{Fourth_order_TM}), labeled by ``TM''
(because it is depicted in a hexagonal shape reminiscent of the elements
of the game board of ``Terra Mystica''), we may choose $\left(\begin{array}{c}
4\\
2
\end{array}\right)$ vertices for the upper position and flip every of the four vertices:
$\ST=6\cdot2^{4}=96$.

The diagram (\ref{Fourth_order_Citroen}) that looks like a doubled
Citro{\"e}n logo (dC), whose symmetry factor is given - analogous
to its third-order-counterpart - by $\SdC=2^{3}=8$. Including the
numerical factors contained in the third-oder-cumulants into the prefactor,
we end up with the following contribution to the effective action:
\begin{equation}
\frac{1}{16}\underbrace{\frac{8\cdot\left(-2\right)^{2}}{24}}_{=\frac{4}{3}}\sum_{i\neq j}J_{ij}^{2}J_{jk}^{2}\left(1-m_{i}^{2}\right)\left(1-m_{j}^{2}\right)\left(1-3m_{i}^{2}\right)\left(1-3m_{j}^{2}\right).\label{eq:Contr_double_Citroen_final}
\end{equation}

Finally, let us determine the symmetry factors for the ``non-standard''
fourth-order-contributions in (\ref{Fourth_order_extraordinary}),
$S_{i}$ for $i\in\left\{ a,b,c\right\} $: For diagram $a$, which
we term ``glasses''-diagram for obvious reasons, there are three
ways to pair, say, the first vertex with any of the three others.
The other pair is then fixed, too. Then, every flip of a vertex produces
a new labeled diagram because it matters which vertex is attached
to the four-point-cumulant. Together, this gives $S_{a}=3\cdot2^{4}$.

For diagram $S_{b}$, flipping each of the two single vertices in
the left part produces a new labeled diagram, because it matters which
vertex is attached to the three-point-cumulant ($2^{2}$). The leaf
brings in a factor $2$, because also here, flipping both vertices
brings in a factor $2$ each and a factor $1/2$ because the leaf
is of second order in the interaction.  We count the leaf as a different
interaction type, therefore the prefactor is given by $\frac{2^{3}}{2!\cdot1!}$.

Finally, there is only one way to construct diagram $c$ and we are
left with the intrinsic symmetry factors of the clusters: $S_{c}=2^{2}$.
Here again, $\Gamma_{V}^{\left(1\right)}$ acts as a special kind
of interaction, which occurs twice, which leads to the prefactor $\frac{2^{2}}{2}$.
Adding up the contributions of the three diagrams leads us to the
following expression for the term inside the square bracket in (\ref{Fourth_order_extraordinary}):\begin{eqnarray}
	&\mkern-95mu\left\{ \frac{1}{4!}3\cdot2^{4}\left(-2\right)\left(1-3m_{j}^{2}\right)+\left(-\frac{1}{2!1!}8+\frac{1}{2}4\right)4m_{j}^{2}\right\} \left(1-m_{i}^{2}\right)\left(1-m_{j}^{2}\right)\left(1-m_{k}^{2}\right)\nonumber\\
	\mkern-115mu=&\mkern-95mu\left\{ -4\left(1-m_{j}^{2}\right)\right\} \left(1-m_{i}^{2}\right)\left(1-m_{j}^{2}\right)\left(1-m_{k}^{2}\right).\\ \nonumber
\end{eqnarray}In conclusion, the diagrams in (\ref{Fourth_order_extraordinary})
add up to
\begin{equation}
\frac{-4}{2^{4}}\sum_{i\neq j\neq k}J_{ij}^{2}J_{jk}^{2}\left\{ 1-m_{j}^{2}\right\} \left(1-m_{i}^{2}\right)\left(1-m_{j}^{2}\right)\left(1-m_{k}^{2}\right).\label{eq:Fourth_order_extraord_raw}
\end{equation}
The term in curly braces, despite emerging from a sum of three different
diagrams, looks like a second cumulant. Interestingly, the contribution
\prettyref{eq:Fourth_order_extraord_raw} for the case $i\neq k$
is exactly canceled by that of the ring-diagram in the case that the
indices of exactly two of the opposite cumulants in the ring are equal.
We see this as follows: There are two ways how exactly two indices
of opposite circles in the ring can be identified ($i=k$ or $j=l$)
so that the symmetry factor of this diagram is multiplied by $2$,
which gives the total prefactor $\frac{1}{4!}\frac{1}{2^{4}}\SR\cdot2=\frac{1}{4}$,
which is indeed equal to the prefactor in \prettyref{eq:Fourth_order_extraord_raw}
(up to the sign). Similarly, we get a contribution to the case of
only two different indices in addition to the ``double Citro{\"e}n-diagram''.
For the diagrams $S_{a}$, $S_{b}$, and $S_{c}$ that means that
the indices of the outer circles are equal, giving the contribution
\begin{equation}
-\frac{1}{4}\sum_{i\neq j}J_{ij}^{2}J_{jk}^{2}\left(1-m_{i}^{2}\right)^{2}\left(1-m_{j}^{2}\right)^{2}.\label{eq:Fourth_order_extraord_outer_equal}
\end{equation}
For the ring diagram, there is a unique way to identify indices of
opposite circles, therefore we do not get the factor $2$ in addition
to the symmetry factor such that this contribution does not cancel
\prettyref{eq:Fourth_order_extraord_outer_equal}, but just reduces
it to $-\frac{1}{16}2\sum_{i\neq j}J_{ij}^{2}J_{jk}^{2}\left(1-m_{i}^{2}\right)^{2}\left(1-m_{j}^{2}\right)^{2}$.
The ``quadruple Citro{\"e}n-diagram'' (\ref{Fourth_order_Citroen})
gives us \prettyref{eq:Contr_double_Citroen_final}. Adding up the
relevant terms of these two contributions, leaving out common factors
and the summation, gives\begin{eqnarray}
&\mkern-65mu-2\left(1-m_{i}^{2}\right)^{2}\left(1-m_{j}^{2}\right)^{2}+\frac{4}{3}\left(1-m_{i}^{2}\right)\left(1-m_{j}^{2}\right)\left(1-3m_{i}^{2}\right)\left(1-3m_{j}^{2}\right)\nonumber\\
\mkern-85mu=&\mkern-65mu\frac{2}{3}\left(1-m_{i}^{2}\right)\left(1-m_{j}^{2}\right)\left(-3\left(1-m_{i}^{2}\right)\left(1-m_{j}^{2}\right)+2\left(1-3m_{i}^{2}\right)\left(1-3m_{j}^{2}\right)\right)\nonumber\\
\mkern-85mu=&\mkern-65mu\frac{2}{3}\left(1-m_{i}^{2}\right)\left(1-m_{j}^{2}\right)\left(-1-3m_{i}^{2}-3m_{j}^{2}+15m_{i}^{2}m_{j}^{2}\right).\\ \nonumber
\end{eqnarray}Taken together this leads to \prettyref{eq:Effective_action_Ising}.

\section{Acknowledgements}

We are indebted to Adam Ran\c con who commented on an earlier version
of this manuscript and especially for his hints on Georges' and Yedidia's
summation convention, which allowed us to see that our result is consistent
with earlier works. Furthermore, we would like to acknowledge the
two anonymous reviewers for many thoughtful comments that helped us
to considerably improve the manuscript. In particular, we thank them
for pointing out the importance of a non-iterative algorithm, that
lead to the idea to use skeleton diagrams, and for pointing us towards
literature on spherical models and laser physics as a potential field
of application.

\providecommand{\newblock}{}

\end{document}